\newif\ifconfver
\confvertrue       
\ifconfver
\documentclass[10pt,twocolumn,twoside]{IEEEtran}
\else
\documentclass[11pt,draftcls,onecolumn]{IEEEtran}
\fi
\usepackage{makecell}
\usepackage{bbding}
\usepackage{color,graphicx}
\usepackage{float}
\usepackage{multirow,amsmath,epsfig,amsfonts,amsbsy,amssymb,psfig,graphics,psfrag,theorem,calc,url,bm,cite}
\usepackage{fancyhdr}
\usepackage{stfloats,hyperref}
\usepackage{algorithm,algorithmic}
\usepackage{diagbox} 
\usepackage{hhline}
\usepackage{subfigure}
\usepackage{comment}
\usepackage{tikz}
\usepackage{quantikz}
\usepackage{pdfrender}
\newtheorem{theorem}{Theorem}
\usepackage{mathtools}

\def\black{\color{black}}

\makeatletter
\def\multilimits@{\bgroup
	\Let@
	\restore@math@cr
	\default@tag
	\baselineskip\fontdimen10 \scriptfont\tw@
	\advance\baselineskip\fontdimen12 \scriptfont\tw@
	\lineskip\thr@@\fontdimen8 \scriptfont\thr@@
	\lineskiplimit\lineskip
	\vbox\bgroup\ialign\bgroup\hfil$\m@th\scriptstyle{##}$\hfil\crcr}
\def\Sb{_\multilimits@}
\def\endSb{\crcr\egroup\egroup\egroup}
\makeatother

{\begin{list}{}%
		{\setlength{\rightmargin}{0pc}%
			\setlength{\leftmargin}{1pc} }} %
	{\end{list}}

\newlength{\twidth}
\ifconfver
\else
\fi
\definecolor{orange}{RGB}{255,107,0}
\def\black{\color{black}}

\usepackage{nameref}

\theorembodyfont{\rmfamily}

{\begin{list}{}{
			\settowidth{\labelwidth}{\mbox{\textnormal{#1}}}%
			\setlength{\leftmargin}{\labelwidth+\labelsep}}}%
	{\end{list}}

\newcommand\bA{\ensuremath{{\bm A}}}
\newcommand\bB{\ensuremath{{\bm B}}}
\newcommand\bC{\ensuremath{{\bm C}}}
\newcommand\bD{\ensuremath{{\bm D}}}

\newcommand\bF{\ensuremath{{\bm F}}}

\newcommand\bI{\ensuremath{{\bm I}}}
\newcommand\bJ{\ensuremath{{\bm J}}}

\newcommand\bM{\ensuremath{{\bm M}}}
\newcommand\bN{\ensuremath{{\bm N}}}
\newcommand\bO{\ensuremath{{\bm O}}}
\newcommand\bP{\ensuremath{{\bm P}}}

\newcommand\bR{\ensuremath{{\bm R}}}
\newcommand\bS{\ensuremath{{\bm S}}}
\newcommand\bT{\ensuremath{{\bm T}}}

\newcommand\bW{\ensuremath{{\bm W}}}

\newcommand\bZ{\ensuremath{{\bm Z}}}

\newcommand\ba{\ensuremath{{\bm a}}}
\newcommand\bb{\ensuremath{{\bm b}}}
\newcommand\bc{\ensuremath{{\bm c}}}

\newcommand\bw{\ensuremath{{\bm w}}}

\newcommand\bz{\ensuremath{{\bm z}}}

\definecolor{orange}{RGB}{255,107,0}
\def\black{\color{black}}

\ifconfver
\author{Chia-Hsiang Lin,~\IEEEmembership{Senior Member,~IEEE}, and Si-Sheng Young,~\IEEEmembership{Student Member,~IEEE}}

\title{Underdetermined Blind Source Separation via Weighted Simplex Shrinkage Regularization and\\ Quantum Deep Image Prior

  \thanks{This study was supported by the Emerging Young Scholar Program (namely, the 2030 Cross-Generation Young Scholars Program) of National Science and Technology Council (NSTC), Taiwan, under Grant NSTC 114-2628-E-006-002.
  We thank the National Center for Theoretical Sciences (NCTS) and the National Center for High-performance Computing (NCHC) for providing the computing resources.}

    \thanks{\textit{(Chia-Hsiang Lin and Si-Sheng Young contribute equally to this work.) (Corresponding author: Chia-Hsiang Lin.)}}
    
    \thanks{C.-H. Lin is with the Department of Electrical Engineering, National Cheng Kung University, Tainan, Taiwan (R.O.C.) (e-mail: chiahsiang.steven.lin@gmail.com).}
    
    \thanks{S.-S. Young is with the Institute of Computer and Communication Engineering, Department of Electrical Engineering, National Cheng Kung University, Tainan, Taiwan (R.O.C.) (e-mail: q38121509@gs.ncku.edu.tw).}
}

\fancypagestyle{IEEEtitlepagestyle}{
	\fancyhf{}
	\fancyhead[LE,RO]{\thepage}
	\fancyfoot[C]{\scriptsize © 2026 IEEE. Personal use of this material is permitted. Permission from IEEE must be obtained for all other uses, in any current or future media, including reprinting/republishing this material for advertising or promotional purposes, creating new collective works, for resale or redistribution to servers or lists, or reuse of any copyrighted component of this work in other works. This is the accepted version of the manuscript. The final, published version is available at IEEE Xplore via https://doi.org/10.1109/TIP.2026.3673957.}
	
}

\else

\fi

\begin{document}
	
	\bibliographystyle{IEEEtran}
	\maketitle
	\ifconfver \else \vspace{-0.5cm}\fi
	
\begin{abstract}
As most optical satellites remotely acquire multispectral images (MSIs) with limited spatial resolution, multispectral unmixing (MU) becomes a critical signal processing technology for analyzing the pure material spectra for high-precision classification and identification.
Unlike the widely investigated hyperspectral unmixing (HU) problem, MU is much more challenging as it corresponds to the underdetermined blind source separation (BSS) problem, where the number of sources is larger than the number of available multispectral bands.
In this article, we transform MU into its overdetermined counterpart (i.e., HU) by inventing a radically new quantum deep image prior (QDIP), which relies on the virtual band-splitting task conducted on the observed MSI for generating the virtual hyperspectral image (HSI).
Then, we perform HU on the virtual HSI to obtain the virtual hyperspectral sources.
Though HU is overdetermined, it still suffers from the ill-posed issue, for which we employ the convex geometry structure of the HSI pixels to customize a weighted simplex shrinkage (WSS) regularizer to mitigate the ill-posedness.
Finally, the virtual hyperspectral sources are spectrally downsampled to obtain the desired multispectral sources.
The proposed geometry/quantum-empowered MU (GQ-$\mu$) algorithm can also effectively obtain the spatial abundance distribution map for each source, where the geometric WSS regularization is adaptively and automatically controlled based on the sparsity pattern of the abundance tensor.
Simulation and {\black real-world data} experiments demonstrate the practicality of our unsupervised GQ-$\mu$ algorithm for the challenging MU task.
Ablation study demonstrates the strength of QDIP, not achieved by classical DIP, and validates the mechanics-inspired WSS geometry regularizer.
The associated code will be available at \url{https://github.com/IHCLab/GQ-mu}. 
The Supplementary Material can be found at \href{https://drive.google.com/file/d/1VzGLL058HbkylZmllq0krSXziaBG_ZXg/view?usp=drive_link}{the provided link}.

%---------
\bfseries{\em Index Terms---}
Multispectral unmixing, 
hyperspectral unmixing, 
multispectral image,
underdetermined blind source separation,
hybrid quantum-classical computing,
quantum deep learning,
deep image prior,
convex geometry.
%---------
\end{abstract}

	\ifconfver \else \vspace{-0.0cm}\fi
	
	\ifconfver \else \vspace{-0.5cm}\fi
	
	%ken does some tricks with line spacing: end here
	\ifconfver \else  \fi

\vspace{-0.2cm}

\section{Introduction} \label{sec: introduction}
Simplex-structured matrix/tensor factorization has {\black emerged as} a critical branch in the blind source separation {\black (BSS)} area due to its wide applicability to solve numerous real-world problems, including the NP-hard spectrum unmixing \cite{MVIE}.
{\black In recent years}, popular optical remote sensing satellites (e.g., Sentinel-2) capture multispectral images (MSIs) over different landscapes, enabling various crucial applications \cite{PWTdehazing}.
However, most optical satellites remotely acquire MSIs with limited spatial resolution; {\black therefore}, multispectral unmixing (MU) becomes a critical signal processing {\black technique} to resolve the so-called mixed pixel phenomenon \cite{HU_TIP,TGFA-AD}.
{\black Although} the MU technique {\black can facilitate} the analysis of pure material spectra for high-precision material classification and identification, it {\black has been} rarely investigated {\black compared} to the widely studied hyperspectral unmixing (HU) problem \cite{HyperCSI,10103834}.
Both MU and HU are {\black BSS} problems, aiming at separating the mixed pixel into pure endmember spectra (sources), while MU is much more challenging as it corresponds to the underdetermined BSS.
Underdetermined BSS addresses the scenario wherein the number of sources is larger than the number of available multispectral bands.

In this article, {\black since} MU {\black is transformed} into its overdetermined counterpart (i.e., HU), we recall some HU techniques {\black in the following}.
Numerous NP-hard HU criteria (e.g., convex geometry-based criteria \cite{tip_citation_1} and non-negative matrix factorization (NMF)-based criteria \cite{NMFcomplexity}) have been {\black proposed} in {\black existing studies}.
Also, deep neural networks (DNN) have attracted {\black significant} attention recently, so {\black a variety of} DNN-based algorithms have {\black demonstrated} their {\black effectiveness} in the HU task \cite{DAAN,MiSiCNet}.
Existing literature rarely discusses the underdetermined MU problem; the only {\black relevant work} is \cite{MU1}, which tries to solve MU by extending several benchmark HU algorithms, including vertex component analysis (VCA) \cite{VCA}, N-FINDR \cite{N-FINDR}, and NMF \cite{NMF-HU}.
However, the experiments in \cite{MU1} {\black consider} only the overdetermined cases, such as a scenario with $P:=8$ MSI bands and $N:=4$ materials/sources.
In other words, there is no existing literature that fundamentally solves the underdetermined MU challenge, as far as we have observed.
We remark that unlike the algebra-based NMF, both VCA and N-FINDR are geometry-based HU methods, and rely on the so-called pure-pixel assumption (PPA) \cite{VCA}, meaning that each source has at least one pixel exclusively composed of that source.
Considering that PPA is often violated in remote sensing, hyperplane-based Craig-simplex-identification (HyperCSI) \cite{HyperCSI} algorithm has been proposed to conduct fast HU without relying on PPA, and the Craig criterion behind has theoretically guaranteed source identifiability \cite{lin2015identifiability}.
To mitigate the NP-hardness, the polynomial-time conic HU criterion has also been proposed based on the Löwner-John ellipsoid in recent machine learning literature \cite{HISUN}.

Beyond image processing, {\black we briefly review} some underdetermined BSS literature, {\black which} are mostly {\black developed} for speech signals (thus unsuitable for image-type signals).
First, a {\black statistical} approach based on Wigner-Ville distribution (WVD) is proposed for time-frequency signals underdetermined BSS in \cite{Underdetermined_BSS_TNNLS_2}.
The modified mixing matrix estimation is proposed to consider the negative value of the auto WVD of the mixture signal in \cite{Underdetermined_BSS_TNNLS_2}, {\black thereby} effectively resolving the underdetermined system.
Another {\black representative approach} is proposed in \cite{ma2010novel}, where singular spectrum analysis is leveraged for those statistically independent sources.
However, the statistically independent assumption does not seem to be true for optical satellite image signals \cite{nascimento2005does}.

To understand the difficulty of the underdetermined MU, we mathematically {\black formulate} our target problem {\black under the} matrix/tensor factorization {\black framework}.
Given a $P$-band MSI $\mathbf{Z}_m \in\mathbb{R}^{L_1 \times L_2 \times P}$ with $L=L_1 L_2$ pixels, MU {\black seeks} to recover the pure endmember matrix $\bB\in\mathbb{R}^{P \times N}$ and the three-way abundance tensor $\mathbf{S} \in\mathbb{R}^{L_1 \times L_2 \times N}$ from the {\black observed} MSI (i.e., $\mathbf{Z}_m=\mathbf{S}\times_3 \bB$ ), where $N$ and ``$\times_3$" denote the number of sources and the tensor mode-$3$ multiplication [defined in the paragraph {\black following} \eqref{DF_1}], respectively.
In this work, we focus on the underdetermined scenario, i.e., $P<N$.
This scenario is more {\black practically relevant}, because a typical multispectral satellite image has a much smaller number of bands $P$ when {\black compared with} hyperspectral satellite ones.
As mentioned, we will tackle this underdetermined MU problem in its counterpart domain (i.e., HU), where we obtain the hyperspectral sources that are {\black subsequently} downsampled back to the multispectral sources.

Specifically, inspired by the prism embedded within the optical satellite sensor, we develop an effective {\black yet} efficient approach to split the MSI into a virtual HSI $\mathbf{Z}_h \in\mathbb{R}^{L_1 \times L_2 \times M}$, where $M>N$ denotes the number of hyperspectral bands.
The MSI is first spectrally split into the hyperspectral domain by the proposed band-splitting (BSP) function, resulting in an auxiliary HSI-like variable with prism attributes.
Considering the nonlinearity of the real-world prism, we employ an additional deep regularization, leading to {\black an efficient} computing of the virtual HSI.
Remarkably, this fully unsupervised approach is quite different from the existing spectral super-resolution (SSR) methods, because the mapping of MSI into HSI in typical SSR relies on {\black a large} amount of training data \cite{MST++}.
Besides, such mapping is adaptive for different real hyperspectral sensors (e.g., cooled CCD camera used in CAVE datasets \cite{CAVE_0293}) while the approach should be general for any MSIs captured by optical sensors.
Conversely, we just need a virtual HSI to gain the desired overdetermined system, and hence it {\black does not} need to be associated with a real HSI sensor; such a high flexibility {\black enables} a general framework to tackle the underdetermined MU problem.

With the {\black obtained} virtual HSI, we perform HU on it to obtain the virtual hyperspectral sources.
{\black Although} HU is overdetermined, it still suffers from {\black ill-posedness}, for which we employ the convex geometry structure of the HSI data pixels to {\black design} a weighted simplex shrinkage (WSS) regularizer to mitigate this issue.
Specifically, \cite{lin2015identifiability} demonstrates that the true spectral signatures are the vertices of the minimum-volume data-enclosing simplex under a rather mild sufficient condition (i.e., when the data purity $\gamma>1/\sqrt{N-1}$), thereby leading to various computationally efficient quadratic regularizers for estimating the signatures; for example, the ``Center" regularizer in \cite{centerMVES} tries to minimize the distance between each signature and the simplex center.
The proposed WSS regularizer {\black enforces} the signatures to the center but using different weights, which can be adaptively and automatically controlled based on the sparsity pattern of the abundance tensor.
We {\black further} introduce the advanced quantum deep network (QUEEN) technique \cite{HyperKING} rather than classical models to address the challenging abundance estimation task.
{\black Specifically}, the depth of classical networks has been theoretically proven to induce a rank-diminishing issue \cite{feng2022rank}.
Since multispectral endmembers are linearly dependent, which may refer to the same abundance in the low-dimensional space (to be shown in Section \ref{subsec:Evaluations}). 
The classical deep layer selecting task-relevant features in a low-dimensional space \cite{lu2021learning,ansuini2019intrinsic} hinders the abundance estimation in underdetermined cases. 
{\black In contrast}, quantum machine learning (e.g., QUEENs) adopts unitary-computing for quantum feature extraction, hence preserving the rank of input data unaltered (thanks to the full rank of unitary quantum operators) \cite{QEDNet,QUEENG}, and is able to transform the data into higher-dimensional space \cite{caro2022generalization}.
The unitary feature learning significantly contributes to solving challenging inverse problems \cite{QEDNet,QUEENG}.
These advantages motivate us to incorporate QUEENs to tackle the abundance estimation.
More detailed explanation for adopting QUEEN is summarized in Section \ref{subsec:QDIP}.
To ensure a fully unsupervised setting, we will implement QUEEN using the concept of deep image prior (DIP) \cite{DIP2018} that considers the deep network itself as a regularizer.
{\black Consequently}, the aforementioned virtual hyperspectral sources are spectrally downsampled to obtain the desired multispectral sources. 
With a customized MU experimental protocol, the {\black effectiveness} of the proposed geometry/quantum-empowered MU (GQ-$\mu$) algorithm will be demonstrated.

The remaining parts of this paper are organized as below.
The details of the proposed GQ-$\mu$ algorithm, including the criterion design, implementation tricks, as well as the architecture of the quantum DIP (QDIP), will be collectively presented in Section \ref{sec:propoed method}.  
Next, the designed experimental protocol, experimental settings, and evaluations are presented and discussed in Section \ref{sec:Experiments}.
Finally, we conclude this paper in Section \ref{sec:conclusion}.

Some notations are defined {\black as follows}.
$\mathbb{Z}_{++}$ {\black denotes} the set of positive integers.
Given $N\in \mathbb{Z}_{++}$, we define $\mathcal{I}_N\triangleq\{1,\dots,N\}$.
$\mathbb{R}^N$, $\mathbb{R}^{M\times N}$, and $\mathbb{R}^{M\times N \times L}$ are the $N$-dimensional Euclidean space, $(M\times N)$-dimensional real-valued matrix space, and $(M\times N \times L)$-dimensional real-valued 3-way tensor space, respectively.
$\operatorname{conv} \mathcal{S}$ is the convex hull of $\mathcal{S}$.
$\bm 1$ (resp., $\bm 0$) {\black denotes} the all-one vector/matrix (resp., all-zero matrix) with the dimensionality specified in its subscript.
$\bm I_{m}$ is the $m\times m$ identity matrix.
$[\mathbf{T}]_{l_1,l_2,\dots,l_N}$ represents the $(l_i,l_2,\dots,l_N)$th entry of the $N$-way tensor $\mathbf{T}$. 
$\bM\geq0$ and $\mathbf{T}\geq0$ denote non-negative matrix and tensor, respectively.
$\bM^{T}$ and $\bM^{-1}$ {\black denote} the matrix transpose and inverse, respectively.
$\text{vec}(\bM)$ denotes the vector obtained by sequentially stacking the columns $\bM$. 
$|\cdot|$, $\|\cdot\|_2$, and $\|\cdot\|_F$ are the absolute value, $\ell_2$-norm, and Frobenius norm, respectively.
$\otimes$ is the Kronecker product.
$\text{Diag}(w_1,\cdots,w_N)$ denotes diagonal matrix with the $n$th diagonal entry being $w_n,~\forall n\in\mathcal{I}_N$.

\section{The Proposed Multispectral Unmixing Framework}\label{sec:propoed method}

In this section, we {\black describe} how we tackle the challenging underdetermined MU problem.
Our approach is outlined {\black as follows}.
Initially, we propose an unsupervised approach with BSP function and deep regularization for obtaining the virtual HSI (cf. Section \ref{subsec:Virtual HSI}).
With this virtual HSI, we transform the MU problem to its overdetermined counterpart (i.e., HU).
Thus, the data-fitting model in HU can be explicitly written as a non-negative tensor factorization problem for hyperspectral source recovery.
Furthermore, as the HU task still suffers from {\black ill-posedness}, we {\black design} a WSS regularizer to mitigate the ill-posedness.
Also, we incorporate the quantum abundance obtained from unsupervised QDIP into the regularization term to address the abundance estimation task.
Details {\black on} designing the data-fitting and regularization terms are summarized in Section \ref{subsec:Criterion Design}. 
Then, the implementations of the overall MU algorithm (i.e., GQ-$\mu$) are presented in Section \ref{subsec:Closed-form}.
Finally, the quantum architecture of QDIP is presented in Section \ref{subsec:QDIP}.

% \vspace{-0.2cm}

\subsection{Criterion Design}\label{subsec:Criterion Design}
Given a three-way MSI tensor $\mathbf{Z}_m$, it can be factorized into a non-negative endmember matrix $\bB$ and a non-negative three-way abundance tensor $\mathbf{S}$.
Specifically, according to the {\black well-known} linear mixing model (LMM) \cite{HU_TIP}, the $(\ell_1, \ell_2)$th pixel of the MSI $[\mathbf{Z}_m]_{\ell_1, \ell_2, :}$ can be {\black expressed as} a specific linear combination of the multispectral endmembers, i.e., 
\[
[\mathbf{Z}_m]_{\ell_1,\ell_2,:}
=\sum_{n=1}^N [\mathbf{S}]_{\ell_1,\ell_2,n}~\!\bb_n,~\forall (\ell_1,\ell_2)\in\mathcal{I}_{L_1}\times\mathcal{I}_{L_2},
\]
where $\bb_n\in\mathbb{R}^{P}$ {\black denotes} the $n$th multispectral endmember (i.e., the $n$th column of $\bB$).
We assume that the model-order $N$ is known a prior for MU or HU \cite{lin2017MDL}.  
{\black Thus}, a natural MU criterion is to minimize the following data-fitting {\black objective},
\begin{align} \label{DF_1}
\textrm{DF}(\bB,\mathbf{S})
=
\frac{1}{2}\|\mathbf{Z}_m-\mathbf{S}\times_3 \bB\|^2_F,
\end{align}
where $\times_3$ denotes the tensor mode-$3$ multiplication.
{\black More precisely}, for a tensor $\mathbf{T}$ and a matrix $\bM$, the tensor mode-$n$ multiplication (denoted by $\mathbf{T} \times_n \bM$) is to left-multiply $\bM$ to each vector along the dimension $n$ within $\mathbf{T}$ (i.e., $[\mathbf{T}\times_n\bM]_{i_1,\dots,i_{n-1},j,i_{n+1},\dots,i_N}\triangleq\sum_m\bM_{j,m}[\mathbf{T}]_{i_1,\dots,i_{n-1},m,i_{n+1},\dots,i_N}$).
{\black It can be observed} that \eqref{DF_1} refers to an underdetermined system in typical MU ($P<N$), so directly solving \eqref{DF_1} is ineffective.
A natural idea for solving the underdetermined issue is to gain more observations/equations, probably virtually.
The idea is inspired by the recent advances of the spectral super-resolution (SSR) technologies \cite{ssr_tip,NatureCom} that map the MSI into its hyperspectral counterpart, and we will show that the idea is feasible. 

According to a typical SSR model, the relation between a MSI and its $M$-band hyperspectral counterpart $\mathbf{Z}_h\in\mathbb{R}^{L_1\times L_2 \times M}$ can be {\black modeled} as $\mathbf{Z}_m=\mathbf{Z}_h\times_3 \bD$ with a spectral response matrix $\bD\in\mathbb{R}^{P \times M}$ \cite{ssr_tip}.
{\black Moreover}, the hyperspectral version of LMM can be {\black formulated} as $\mathbf{Z}_h=\mathbf{S} \times_3 \bA$, where $\bA\triangleq[\ba_1,\dots,\ba_N]\in\mathbb{R}^{M\times N}$ with $\ba_n$ being the $n$th hyperspectral endmember.
{\black Thus}, we have $\mathbf{Z}_m=\mathbf{S} \times_3 \bA \times_3\bD$, and {\black hence} \eqref{DF_1} can be {\black equivalently} reformulated as
\begin{align} \label{DF_2}
\frac{1}{2}\|\mathbf{Z}_m-\mathbf{S}\times_3 \bB\|^2_F=\frac{1}{2}\|\mathbf{Z}_m-\mathbf{S}\times_3 \bA \times_3 \bD\|^2_F,
\end{align}
which {\black enables} us to estimate the hyperspectral signatures $\bA$ (overdetermined) rather than the multispectral ones $\bB$ (underdetermined).
{\black Accordingly}, we incorporate the HSI $\mathbf{Z}_h$ into this underdetermined system, thereby {\black yielding} a new data-fitting model for the MU task, i.e.
\begin{equation} \label{DF_term}
\text{DF}(\bA,\mathbf{S})\!\triangleq\!\frac{1}{2}\|\mathbf{Z}_m\!-\!\mathbf{S}\!\times_3\! \bA \!\times_3 \!\bD\|^2_F\!+\!\frac{1}{2}\|\mathbf{Z}_h\!-\!\mathbf{S}\! \times_3 \!\bA\|^2_F.
\end{equation}
We {\black note} that \eqref{DF_term} is similar to a typical unmixing-based fusion problem \cite{COCNMF}, yet it is {\black fundamentally} different. 
Specifically, the MSI, HSI, spectral response, and spatial downsampling matrix are {\black typically} required for data fusion, while our MU framework requires only MSI as the input.
The HSI $\mathbf{Z}_h$ here is virtual, which is obtained by prism-inspired BSP function and deep regularization (cf. Section \ref{subsec:Virtual HSI}), rather than real hyperspectral sensors.
This design {\black allows} us to characterize their relations using a virtual $\bD$ without considering resolution variability between $\mathbf{Z}_m$ and $\mathbf{Z}_h$.
Thus, $\mathbf{Z}_m$ and $\mathbf{Z}_h$ in the virtual framework are spatially-aligned and necessarily share the same abundances.
Besides, the virtuality of $\mathbf{Z}_h$ directly provides a derivable $\bD$ (directly determined by BSP) without additional estimation, while the spectral response between two real sensors is typically unavailable (requiring estimation).
Conversely, if we already have a real hyperspectral sensor with a real $\bD$, why would we still need to address the MU problem? 
In this case, we just have an overdetermined system, {\black with} no need to solve the underdetermined MU problem.
Thus, adopting a virtual $\bD$ is actually a natural approach for the MU problem.
This practical limitation (unavailability of real hyperspectral sensors) also makes the typical data-driven SSR methods invalid for our target task.
Hence, we obtain $\mathbf{Z}_h$ by the unsupervised spectral augmentation method, resulting in a highly flexible, general, and unsupervised SSR stage, as detailed in Section \ref{subsec:Virtual HSI}.
Through \eqref{DF_term}, we are able to explore the spatial characteristics from $\mathbf{Z}_m$ and spectral attributes from $\mathbf{Z}_h$, thereby solving the challenging underdetermined MU.

Nevertheless, the data-fitting model alone still suffers from the ill-posed issue, for which we {\black design several} regularization schemes to mitigate the ill-posedness of \eqref{DF_term}.
First, based on the {\black intrinsic properties} of remote sensing images, a specific material is usually distributed only around certain regions of the image.
{\black Thus}, a valid abundance tensor is often sparse \cite{Sparse_TIP}, resulting in the $\ell_1$-norm regularization, i.e.,
\begin{align} \label{l1-norm REG}
\lambda_1\| \mathbf{S}\|_1\triangleq  \lambda_1\sum_{n=1}^{N} \sum_{\ell_2=1}^{L_2} \sum_{\ell_1=1}^{L_1} |[\mathbf{S}]_{\ell_1,\ell_2,n}|,
\end{align}
where $\lambda_1\geq 0$ controls the regularization strength.
However, {\black relying solely} on the sparsity feature is not sufficient to regularize the challenging problem.
{\black Therefore}, we {\black further} explore quantum features for the regularization of $\mathbf{S}$.
Once the quantum abundance feature $\mathbf{S}_{\text{qu}}\in\mathbb{R}^{L_1\times L_2 \times N}$ is extracted via QDIP (to be done in Section \ref{subsec:QDIP}), its {\black information} is {\black incorporated} into the final abundance solution $\mathbf{S}$ by minimizing this regularizer,
\begin{align} \label{Q-norm REG}
\frac{\lambda_2}{2}\| \mathbf{S}-\mathbf{S}_{\text{qu}}\|_F^2,
\end{align}
where $\lambda_2\geq 0$ controls the regularization strength.
As {\black discussed} in Section \ref{subsec:QDIP}, the quantum abundance feature $\mathbf{S}_{\text{qu}}$ will be computed by QUEEN (rather than traditional deep networks, such as convolution neural network (CNN)), because QUEEN can extract unique quantum unitary-computing features using very light-weight network architecture (thanks to the quantum full expressibility \cite[Theorem 2]{HyperQUEEN}).
Notably, for the first time, the unsupervised version of QUEEN is employed to solve the matrix/tensor factorization problem, and experimental results evidence the effectiveness of our customized quantum regularization (cf. Section \ref{subsec:Ablation Study}).

\begin{figure}[t]
    \centering
    \includegraphics[width=1\linewidth]{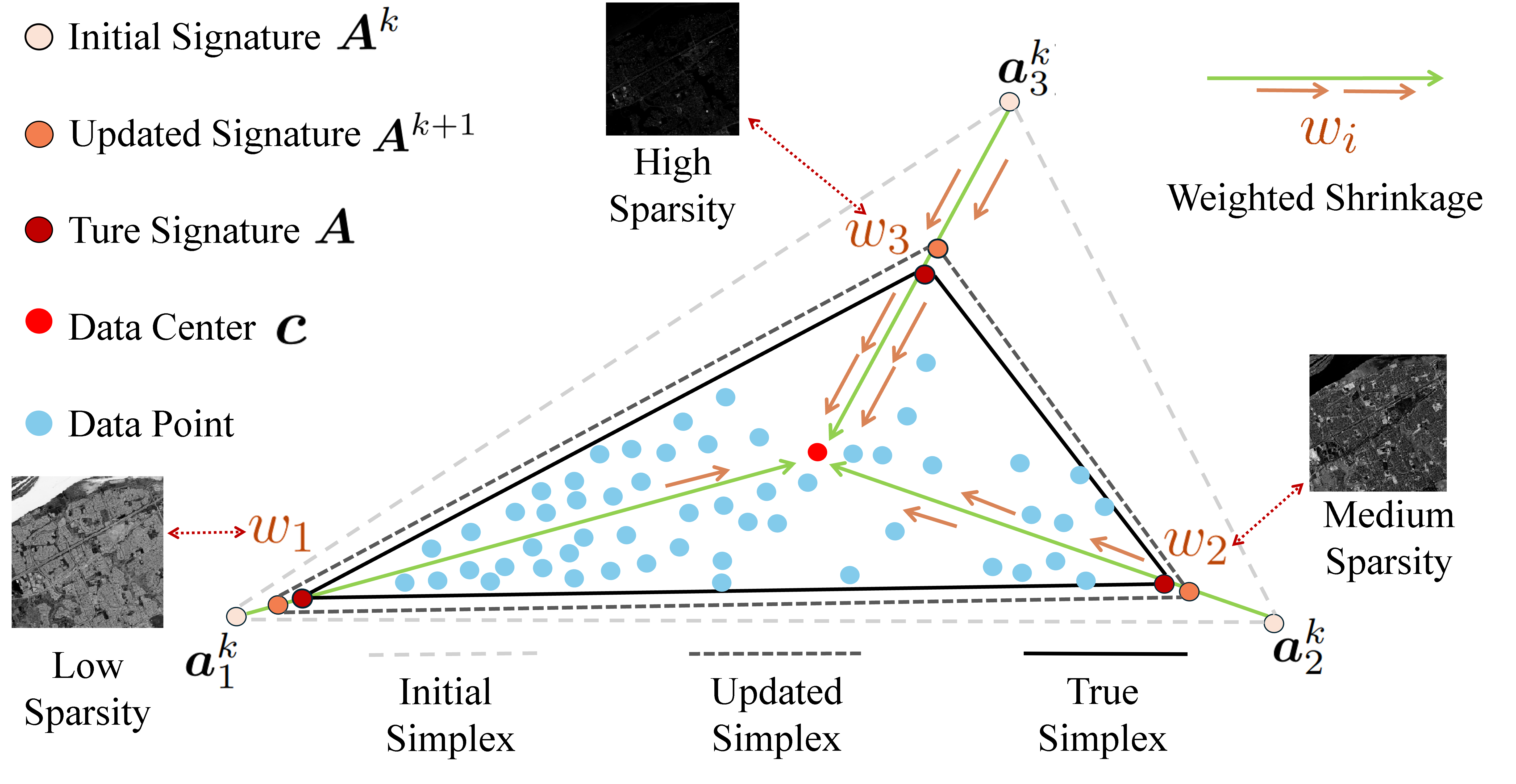}
    \caption{Graphical illustration of the proposed weighted simplex shrinkage (WSS) regularization.
    Unlike existing volume-minimization types of regularization that blindly promote the volume minimization without considering the sparsity pattern of the abundances, the proposed WSS regularization further takes the sparsity level into consideration.
    Intuitively, when an endmember $\ba_n^{k}$ corresponds to a highly sparse abundance map at the $k$th algorithmic iteration, this endmember is geometrically very far away from the data center $\bc$.
    So, the endmember requires a larger force (i.e., a larger weight $w_3$) to draw it to be closer to $\bc$ at the $(k+1)$th iteration.
    By contrast, for the relatively dense abundance map (low sparsity), the corresponding endmember just needs to be slightly pushed to the center $\bc$, thus requiring a smaller weight (e.g., $w_1$).
    A larger weight $w_n$ will be associated with more orange arrows (denoting a higher force).
    }
    \label{fig: simplex_shrinkage}
\end{figure}

Furthermore, we propose a geometry-based WSS {\black regularizer} for endmember matrix estimation.
It is known that the hyperspectral endmember matrix $\bA=[\ba_1, \cdots, \ba_N]$ {\black can} be perfectly recovered from the vertices of the minimum-volume data-enclosing simplex $\operatorname{conv} \{\ba_1, \cdots, \ba_N\}$, {\black under} a very mild sufficient condition (i.e., when the data purity $\gamma>1/\sqrt{N-1}$) \cite{lin2015identifiability}.
Therefore, numerous regularization schemes {\black have been} proposed to minimize the volume of the {\black corresponding} endmember simplex $\operatorname{conv} \{\ba_1, \cdots, \ba_N\}$.
Mathematically, the simplex volume can be {\black formulated} as $\frac{1}{(N-1)!} \left( \det(\bar{A}^{T}\bar{A}) \right)^{\frac{1}{2}}$, {\black where} $\bar{\mathbf{A}} \triangleq \left[\ba_1 - \ba_N, \cdots, \ba_{N-1} - \mathbf{a}_N \right]$ \cite{COCNMF}.
However, the inherent non-convexity of the determinant-based simplex volume regularization {\black may} complicate the gradient descent procedure in NMF frameworks \cite{MVCNMF}.   
As {\black an alternative}, existing literature adopts convex {\black surrogates}, sum-of-squared distance (SSD) between the endmembers, i.e., $\psi(\bA)=\frac{1}{2}\sum_{i=1}^{N-1}\sum_{j=i+1}^{N}\| \ba_i-\ba_j\|_2^2$ \cite{COCNMF}, {\black which} enables an efficient implementation.
In addition, quadratic minimum simplex volume regularizations \cite{CoNMF,centerMVES} hold the general form,
\begin{align} \label{QMV}
\frac{1}{2}\| \bA\bZ- \bO\|_F^2,
\end{align}
where $\bZ\in\mathbb{R}^{N \times N}$ and $\bO\in\mathbb{R}^{M \times N}$ are some specific matrices {\black corresponding to} different surrogates, including ``Center'', ``Boundary'', and ``TV''.
For {\black example}, the ``Center'' regularization refers to $(\bZ,\bO)\triangleq(\bI_N,\bc\mathbf{1}_N^T)$ with $\bc\triangleq \frac{1}{L}\sum_{(\ell_1,\ell_2)} [\mathbf{Z}_h]_{\ell_1,\ell_2,:}$, which minimizes the simplex volume by shrinking the vertices (i.e., the columns of $\bA$) toward the data center $\bc$ (cf. Figure \ref{fig: simplex_shrinkage}).
{\black Moreover}, when $(\bZ,\bO)\triangleq(\bI_N-\frac{1}{N}\mathbf{1}_{N\times N},\mathbf{0}_{M\times N})$, the surrogate \eqref{QMV} shrinks the simplex volume by minimizing the total variations between vertices, i.e., ``TV'' regularization.
However, these regularizations promote volume minimization without considering the shrinkage ``force'' and ``direction'', potentially resulting in less effective estimations.
Conversely, we {\black design} a new mechanics-inspired WSS geometry regularizer to address these limitations.
First, we conceptually explain the {\black effect} of the ``force''.
Intuitively, when an endmember $\ba_n^{k}$ corresponds to a highly sparse abundance map at the $k$th algorithmic iteration, the endmember is {\black geometrically far} away from the data center $\bc$.
{\black Thus}, this endmember should be forced more to the center at the $(k+1)$th algorithmic iteration, as illustrated in Figure \ref{fig: simplex_shrinkage}.
{\black In contrast}, for a relatively dense abundance map, the corresponding endmember {\black only} needs to be slightly pushed to the center $\bc$ (cf. Figure \ref{fig: simplex_shrinkage}).
{\black Moreover}, it is essential to consider the shrinkage directions.
When the number of algorithmic iterations is large enough, $\ba_n^k$ is {\black expected} to recover the $n$th true simplex vertex.
However, in a typical NMF problem, the endmember and abundance matrices are not jointly convex, {\black which implies} that the optimal endmember and abundance may not be simultaneously achieved.
For instance, when the endmember matrix is well recovered at some iteration $k$ (but not the abundance matrix), the simplex formed by the endmembers tends to be {\black excessively} shrunk by volume regularizers when further optimizing the abundances in the upcoming algorithmic iterations.
Hence, the mechanics-inspired WSS regularization is proposed to simultaneously take the force and direction into account.
We will experimentally {\black demonstrate} that the strategy of simultaneously considering force and direction (formulated below) does lead to a stronger regularization for the challenging MU task in Section \ref{subsec:Ablation Study}.

{\black In practice}, we prefer to implement the {\black aforementioned} idea using a convex function like \eqref{QMV}.
{\black Thus}, we define the weighted simplex shrinkage (WSS) regularizer as
\begin{align}\label{WSS}
    &\text{WSS}(\bA;\bw)\notag
    \\
    &~\triangleq\frac{\lambda_3}{2}\|(\bA-\bc\mathbf{1}_N^T)\sqrt{\text{Diag}(\bw)}\|^2_F
    +\frac{\lambda_4}{2}\|\bA-\bA^{k}\|^2_F,
\end{align}
where $\lambda_3\geq 0$ and $\lambda_4\geq 0$ control the {\black corresponding} regularization strengths, $\bA^{k}$ {\black denotes} the updated endmember matrix at the $k$th iteration, and $\bc:=\frac{1}{N}\sum_{i=1}^N\ba^0_i$ (rather than the mean of all hyperspectral pixels, thereby avoiding potential outlier {\black effects}).
{\black Moreover}, the weighting vector is defined as $\bw\triangleq \text{softmax}(L_1',\dots,L_N')$, {\black where} $L_n'$ {\black denotes} the normalized sparsity level for the $n$th endmember (defined next), and $\text{softmax}(\cdot)$ is the softmax operator \cite{TGFA-AD}.
The sparsity level for the $n$th endmember can be naturally defined as the reciprocal $L_n:=\| [\mathbf{S}^{0}]_{:,:,n}\|_1^{-1}=(\sum_{\ell_2=1}^{L_2} \sum_{\ell_1=1}^{L_1} |[\mathbf{S^0}]_{\ell_1,\ell_2,n}|)^{-1}$, where $\mathbf{S^0}$ is the initial abundance estimation (cf. Section \ref{subsec:Closed-form}).
{\black Subsequently}, the normalized sparsity level is given by $L_n':=L_n/\max\{L_1,\dots,L_N\}$.
To have an in-depth understanding of our WSS regularization design, one {\black may} consider the gradient of our WSS regularization in the case of $\lambda_3:=1$ and $\lambda_4:=1$, which can be explicitly {\black derived} as, 
\begin{align}\label{gradient of WSS}
    \nabla_{\bA} \text{WSS}(\bA;\bw)=(\bA-\bc\mathbf{1}_N^T)\text{Diag}(\bw)+(\bA-\bA^{k}).
\end{align}
For the ``force'', the first term in the RHS of \eqref{gradient of WSS} adaptively shrinks the endmembers to the center based on the sparse pattern $\bw$ of their corresponding abundances.
A higher sparsity level leads to a larger $w_n$ (i.e., the $n$th entry of $\bw$), and hence the first term of \eqref{WSS} will adaptively force $\ba_n$ to the center $\bc$, as desired (cf. Figure \ref{fig: simplex_shrinkage}).
Additionally, the second term in the RHS of \eqref{gradient of WSS} is responsible for regularizing the ``direction'' to prevent the over-shrinkage or biased issues.
Therefore, the second term of \eqref{WSS} encourages a more stable update of the endmember matrix $\bA$ (not too far away from the previous update $\bA^k$) during the optimization procedure.

{\black In summary}, the proposed regularization strategy is explicitly summarized as
\begin{align*}
\text{REG}(\bA,\mathbf{S})\triangleq\lambda_1\| \mathbf{S}\|_1+\frac{\lambda_2}{2}\| \mathbf{S}-\mathbf{S}_{\text{qu}}\|_F^2+\text{WSS}(\bA;\bw),
\end{align*}
resulting in the overall criterion, i.e.,
\begin{align}\label{eq:NTF criterion}
\min_{\bA\geq\mathbf{0},~\mathbf{S}\geq\mathbf{0}}~&~\text{DF}(\bA,\mathbf{S})+\text{REG}(\bA,\mathbf{S}),
\end{align}
in which the two non-negative constraints are to capture the natural properties of the endmembers and abundances.
Once the criterion \eqref{eq:NTF criterion} is solved, its optimal solutions $(\bA^\star,\mathbf{S}^\star)$ directly yields the MU solutions, which are the multispectral endmember matrix $\bB^\star=\bD\bA^\star$ (multispectral sources) and the abundance tensor  $\mathbf{S}^\star$.
The exhaustive implementations of \eqref{eq:NTF criterion} are summarized in Section \ref{subsec:Closed-form}.
Besides, the details for computing the virtual HSI $\mathbf{Z}_h$ and for designing the QDIP are presented in Section \ref{subsec:Virtual HSI} and Section \ref{subsec:QDIP}, respectively.

\subsection{Algorithm Implementation}\label{subsec:Closed-form}
In this section, we {\black present} the GQ-$\mu$ algorithm (i.e., Algorithm \ref{alg:MUmain}) {\black to implement} the criterion \eqref{eq:NTF criterion}.
In the MU task, we pay more attention to the abundance tensor $\mathbf{S}$ because the abundance explicitly {\black characterizes} the distribution of the pure materials.
{\black Thus}, we first update $\mathbf{S}^{k+1}$, and then $\mathbf{A}^{k+1}$.
To this {\black purpose}, we initialize $\bA^0$ and $\mathbf{S}^0$ by performing HU on the virtual HSI $\mathbf{Z}_h$ using the highly-mixed/ill-conditioned spectrum unmixing (HISUN) algorithm \cite{HISUN}, where $\mathbf{Z}_h$ will be computed in Section \ref{subsec:Virtual HSI}.
{\black Subsequently}, we use $\bA^0$, $\mathbf{Z}_h$, and the proposed QDIP to obtain the quantum abundance tensor (QAT) $\mathbf{S}_{\text{qu}}$, and this will be detailed in Section \ref{subsec:QDIP}.
Once we have $\mathbf{S}_{\text{qu}}$, we can then update the $\mathbf{S}^{k+1}$ by solving the first subproblem induced from \eqref{eq:NTF criterion}, i.e., the subproblem for optimizing $\mathbf{S}$ (with fixed $\bA$), which can be derived {\black based on} \eqref{DF_term}, \eqref{l1-norm REG}, \eqref{Q-norm REG} and \eqref{eq:NTF criterion} as
\begin{align}\label{subproblem S} 
\mathbf{S}^{k+1}\!\in \arg~&\min_{\mathbf{S}\geq0}\text{DF}(\bA^{k},\mathbf{S})\!+\!\lambda_1\| \mathbf{S}\|_1
\!+\!\frac{\lambda_2}{2}\| \mathbf{S}\!-\!\mathbf{S}_{\text{qu}}\|_F^2,
\end{align}
which will be optimized using Algorithm \ref{alg:eq:subproblem ADMM} (to be introduced next). 
After we have $\mathbf{S}^{k+1}$, we subsequently update $\bA^{k+1}$ by solving the second subproblem induced from \eqref{eq:NTF criterion}, i.e., the subproblem for optimizing $\bA$ (with fixed $\mathbf{S}$), which can be derived from \eqref{DF_term}, \eqref{WSS} and \eqref{eq:NTF criterion} as
\begin{equation} \label{subproblem A} 
\bA^{k+1}\!\in \arg~\min_{\bA\geq0}\text{DF}(\bA,\mathbf{S}^{k+1})\!+\!\text{WSS}(\bA;\bw).
\end{equation}
Once the above altering optimization converges to $(\bA,\mathbf{S})$, we have the multispectral endmember matrix $\bB^{\star}:=\bD\bA^{\star}$ and the abundance tensor $\mathbf{S}^{\star}$ as the final MU solution, as summarized in Algorithm \ref{alg:MUmain}.

%====================================================
	\begin{algorithm}[t]
		\caption{GQ-$\mu$ Algorithm for Solving \eqref{eq:NTF criterion}}
		%----------------------------------------------------
		\begin{algorithmic}[1]\label{alg:MUmain}
            \STATE Given the MSI $\mathbf{Z}_m$.
            \STATE Construct the virtual HSI $\mathbf{Z}_h$ and spectral response matrix $\bD$ from $\mathbf{Z}_m$ (cf. Section \ref{subsec:Virtual HSI}). Set $k:=0$.
            \STATE Initialize $\bA^k$ and $\mathbf{S}^k$ by conducting HU on $\mathbf{Z}_h$ using HISUN \cite{HISUN}.
            \STATE Extract QAT $\mathbf{S}_{\text{qu}}$ from $\mathbf{Z}_h$ and $\bA^k$ via QDIP (cf. Section \ref{subsec:QDIP}).
            
            \REPEAT

            \STATE Update $\mathbf{S}^{k+1}$ by Algotirhm \ref{alg:eq:subproblem ADMM}.%by \eqref{subproblem S} using using $\bD,\bA^t, \mathbf{S}^t, \mathbf{S}_{\text{qu}}, \mathbf{Z}_m$, and $\mathbf{Z}_h$.
            
            \STATE Update $\bA^{k+1}$ by \eqref{update A implementation} using $\bD,\bA^k,\mathbf{S}^{k+1}, \mathbf{Z}_m$, and $\mathbf{Z}_h$.%\eqref{subproblem A} using $\bD,\bA^t, \mathbf{S}^{t+1}, \mathbf{Z}_m$, and $\mathbf{Z}_h$.
            
		  \STATE $k:=k+1$.
            \UNTIL the predefined stopping criterion is met.
            \STATE {\bf Output} multispectral endmember matrix $\bB^{\star}:=\bD\bA^k$ and abundance tensor $\mathbf{S}^{\star}:=\mathbf{S}^k$.
		\end{algorithmic}
		%----------------------------------------------------
	\end{algorithm}
%====================================================
%
%====================================================
	\begin{algorithm}[t]
		\caption{ADMM Algorithm for Solving \eqref{reformulation S}}
		%----------------------------------------------------
		\begin{algorithmic}[1]\label{alg:eq:subproblem ADMM}
            \STATE Given $\mathbf{Z}_m$, $\mathbf{Z}_h$, $\mathbf{S}_{\text{qu}}$, $\bD$, $\mathbf{S}^k$, and $\bA^k$. Set $q:=0$.
            \STATE Initialize $\mathbf{Y}^{q}:=\mathbf{S}^k$ and $\mathbf{V}^{q}:=\mathbf{0}_{L_1\times L_2\times N}$
            
            \REPEAT
            \STATE Update $\mathbf{S}^{q+1}\in\arg\min_{\mathbf{S}}~\mathcal{L}(\mathbf{Y}^{q},\mathbf{S},\mathbf{V}^{q})$.
		  
            \STATE  Update $\mathbf{Y}^{q+1}\in\arg\min_{\mathbf{Y}}~\mathcal{L}(\mathbf{Y},\mathbf{S}^{q+1},\mathbf{V}^{q})$.
            
            \STATE  Update $\mathbf{V}^{q+1}:=\mathbf{V}^{q}-(\mathbf{Y}^{q+1}-\mathbf{S}^{q+1})$.
            
		  \STATE $q:=q+1$.
            \UNTIL the predefined stopping criterion is met.
            \STATE {\bf Output} $\mathbf{S}^{k+1}:=\mathbf{S}^{q}$.
		\end{algorithmic}
		%----------------------------------------------------
	\end{algorithm}
%====================================================

Next, we solve \eqref{subproblem S}, which is now convex, using the powerful convex {\black optimization} solver known as alternating direction method of multipliers (ADMM) \cite{ADMM_Boyd}.
{\black To handle} the non-differentiable $\ell_1$-norm term in \eqref{subproblem S}, we employ an auxiliary variable $\mathbf{Y}$ to reformulate \eqref{subproblem S} {\black as} the standard ADMM form,
\begin{align}\label{reformulation S} 
&\min_{\mathbf{S}}~~\text{DF}(\bA^{k},\mathbf{S})+\lambda_1\| \mathbf{Y}\|_1
+\frac{\lambda_2}{2}\| \mathbf{S}-\mathbf{S}_{\text{qu}}\|_F^2+I_+(\mathbf{S})\notag,
\\
& ~~\text{s.t.}~~\mathbf{Y}=\mathbf{S},
\end{align} 
where $I_+(\cdot)$ is the indicator function of the non-negative orthant defined as
\[
I_+(\mathbf{S})\triangleq 
\begin{cases}
0,  & \text{if $\mathbf{S}\geq 0$}, \\
\infty, & \text{otherwise.}
\end{cases}
\]
According to the ADMM theory, we first write down the augmented Lagrangian of \eqref{reformulation S}, i.e.,
\begin{align} \label{augmented Lagrangian}
%\label{eq:Lag}
\mathcal{L}(\mathbf{Y},\mathbf{S},\mathbf{V})
\nonumber 
&=
\text{DF}(\bA^{k},\mathbf{S})+\lambda_1\|\mathbf{Y}\|_1+\frac{\lambda_2}{2}\|\mathbf{S}-\mathbf{S}_{\text{qu}}\|_F^2\notag
\\
&
+I_+(\mathbf{S})+\frac{\mu}{2}\|\mathbf{Y}-\mathbf{S}-\mathbf{V}\|_F^2,
\end{align}
where $\mu>0$ and $\mathbf{V}\in\mathbb{R}^{L_1\times L_2 \times N}$ are the penalty parameter and the dual variable, respectively.
{\black Owing to} the dual decomposability of ADMM, it allows us to solve the non-differential term of $\mathbf{Y}$ separately, and the standard ADMM optimization procedure \cite{ADMM_Boyd} is summarized in Algorithm \ref{alg:eq:subproblem ADMM}.

To complete Algorithm \ref{alg:eq:subproblem ADMM}, we {\black need to} solve the two {\black resulting} optimization problems presented in Line 4 (for $\mathbf{S}^{q+1}$) and Line 5 (for $\mathbf{Y}^{q+1}$) of Algorithm \ref{alg:eq:subproblem ADMM}.
{\black For updating} $\mathbf{S}^{q+1}$, the non-negative constraint (associated with the indicator function) {\black leads to} a slow optimization procedure, if the global {\black optimum} is pursued.
To efficiently {\black yet} approximately compute $\mathbf{S}^{q+1}$, we first minimize $\mathcal{L}(\mathbf{Y}^{q},\mathbf{S},\mathbf{V}^{q})$ (with fixed $\mathbf{Y}^{q}$ and $\mathbf{V}^{q}$) without considering the {\black indicator} term of $I_{+}$, followed by projecting the minimizer back to the non-negative orthant.
{\black Regarding} the initialization of the auxiliary variable $\mathbf{Y}^0$, we judiciously {\black adopt} the previous-stage abundance tensor (i.e., $\mathbf{S}^k$) as a warm start strategy \cite{CVXbookCLL2016}, resulting in the explicitly updated expression,
\begin{align}\label{update_S}
\mathbf{S}^{q+1}\equiv\mathbf{S}_{(3)}^{q+1}=\Pi((\bR\!+\!(\lambda_2\!+\!\mu)\bI_N)^{-1}(\bP\!+\!\mu\bT^q)),
\end{align}
where $\bR\triangleq (\bA^{k})^T\bD^{T}\bD\bA^{k}+(\bA^{k})^T\bA^{k}$, $\bP\triangleq(\bA^{k})^T\bD^{T}\mathbf{Z}_{m(3)}+(\bA^{k})^T\mathbf{Z}_{h(3)}+\lambda_2\mathbf{S}_{\text{qu}(3)}$, $\bT^q\triangleq\mathbf{Y}_{(3)}^q-\mathbf{V}_{(3)}^q$, and $\Pi(\cdot)$ is the projector onto the non-negative orthant.
Here, the subscript ``$(n)$" denotes the tensor mode-$n$ matricization.
Specifically, for the $N$-way tensor $\mathbf{G}\in\mathbb{R}^{M_1\times M_2\times \cdots \times M_N}$, its mode-$n$ matricization is denoted as $\mathbf{G}_{(n)}$, and it maps $[\mathbf{G}]_{\ell_1,\cdots,\ell_N}$ to $[\mathbf{G}_{(n)}]_{\ell_n,j}~(j\triangleq 1+\sum_{t=1,t\neq n}^{N} (\ell_t-1)J_t$ and $J_t\triangleq\prod_{m=1,m\neq n}^{t-1} M_m)$.

To update $\mathbf{Y}^{q+1}$ in Line 5 of Algorithm \ref{alg:eq:subproblem ADMM} (with fixed $\mathbf{S}^{q+1}$ and $\mathbf{V}^{q}$), we employ the renowned ``soft-threshold'' (shrinkage) algorithm \cite{soft-thresholding} to solve the $\ell_1$-norm regularization is defined as
\begin{align}\label{eq: soft-threshold}
\mathbf{Y}^{q+1} & ~= \arg\min_{\mathbf{Y}}~\mathcal{L}(\mathbf{Y},\mathbf{S}^{q+1},\mathbf{V}^{q})\notag
\\
&
 ~=\arg\min_{\mathbf{Y}}~\lambda_1\|\mathbf{Y}\|_1+\frac{\mu}{2}\|\mathbf{Y}-\mathbf{S}^{q+1}-\mathbf{V}^{q}\|_F^2\notag,
 \\
&
 ~=
 \eta_{ {\lambda_1}/{\mu} }(\mathbf{S}^{q+1}+\mathbf{V}^{q}),
\end{align}
where $\eta_c(\cdot)$ denotes the shrinkage operator.
In detail, given a three-way tensor $\mathbf{T}\in\mathbb{R}^{L_1 \times L_2 \times N}$ and a constant $c\geq 0$, the $(\ell_1,\ell_2,n)$th element of the output three-way tensor $\eta_c(\mathbf{T})$ is defined as
\[
[\eta_c(\mathbf{T})]_{\ell_1,\ell_2,n}\triangleq 
\begin{cases}
[\mathbf{T}]_{\ell_1,\ell_2,n}-c,  & \text{if $[\mathbf{T}]_{\ell_1,\ell_2,n}\geq c$}, 
\\
0, & \text{if $|[\mathbf{T}]_{\ell_1,\ell_2,n}|\leq c$}, 
\\
[\mathbf{T}]_{\ell_1,\ell_2,n}+c, & \text{if $[\mathbf{T}]_{\ell_1,\ell_2,n}\leq -c$}.
\end{cases}
\]
The derivation of Algorithm \ref{alg:eq:subproblem ADMM} is thus completed, and {\black consequently} Line 6 of Algorithm \ref{alg:MUmain} is done.

The remaining task is to derive Line 7 of Algorithm \ref{alg:MUmain}, which is to update the endmember matrix $\bA^{k+1}$ by solving \eqref{subproblem A}.
To explicitly derive it, we equivalently reformulate \eqref{subproblem A} as
\begin{align}\label{vector representation A}
&
\min_{\ba\geq 0}~\frac{1}{2}\|\bz_m-(\bS^{T}\otimes\bD)\ba\|^2_2+\frac{1}{2}\|\bz_h-(\bS^T\otimes\bI_M)\ba\|^2_2\notag
\\
&
+
\frac{\lambda_3}{2}\|(\sqrt{\bW}\otimes\bI_M)(\ba-\widehat{\bc})\|^2_2+\frac{\lambda_4}{2}\|\ba-\ba^{k}\|^2_2,
\end{align}
where $\bz_m\triangleq\text{vec}(\mathbf{Z}_{m(3)})$, $\bS\triangleq\mathbf{S}_{(3)}^{(k+1)}$, $\ba\triangleq\text{vec}(\bA)$,  $\bz_h\triangleq\text{vec}(\mathbf{Z}_{h(3)})$, $\bW\triangleq\text{Diag}(\bw)$, $\widehat{\bc}\triangleq\text{vec}(\bc\mathbf{1}_N^T)$, and $\ba^{k}\triangleq\text{vec}(\bA^{k})$.
{\black For} the non-negative constraint of the signature matrix, we adopt the same trick to first ignore it (followed by projecting the solution back to the non-negative orthant), thereby leading to an efficient update of \eqref{vector representation A}, i.e., 
\begin{align}\label{update A implementation}
\!\!\!\bA^{k+1}\!\equiv\!\ba^{k+1}\!=\!\Pi(\bJ_1^{-1}(\bJ_2\bz_m\!+\!\bJ_3\bz_h\!+\!\lambda_3\bJ_4\widehat{\bc}\!+\!\lambda_4\ba^{k})),
\end{align}
where $\bJ_1\triangleq(\bS\bS^T\otimes\bD^T\bD)+ (\bS\bS^T\otimes\bI_M)+\lambda_3\bJ_4+\lambda_4\bI_{NM}$, $\bJ_2\triangleq (\bS\otimes\bD^T)$, $\bJ_3\triangleq(\bS\otimes\bI_M)$, and $\bJ_4\triangleq(\bW\otimes\bI_M)$.
According to the above implementations, we already completed the proposed GQ-$\mu$ algorithm (cf. Algorithm \ref{alg:MUmain}).
\subsection{Spectral Augmentation}\label{subsec:Virtual HSI}
In this section, we {\black describe} the approach for obtaining $\mathbf{Z}_h$.
As {\black emphasized} before (cf. Section \ref{sec: introduction}), this approach should be general for any MSIs with different spectral responses, {\black which implies} that {\black existing} data-driven SSR methods \cite{NatureCom} are unsuitable for our target application.
Accordingly, optimization criteria with sophisticated regularizations are inevitable to ensure that $\mathbf{Z}_h$ {\black preserves} the natural properties (e.g., non-negativity, and spectral-spatial continuity) to be a valid HSI.
To address this {\black challenge}, we develop a simple yet effective approach to obtain the virtual HSI.
Specifically, the natural properties of real-world HSIs {\black originate} from the prism embedded within optical systems, meaning that the HSI can be obtained by further {\black performing} the light splitting on its counterpart MSI. 
Conversely, MSI can be considered as a weighted summation along the spectral dimension of its hyperspectral counterpart.
This relation is the {\black fundamental} concept of the SSR technique.
For example, in an $ M$-time SSR, the green spectral band of an MSI can be approximated by a weighted summation of $M$ hyperspectral bands within the green wavelength range, while these non-negative weights are determined by the spectral response function (SRF).
{\black Accordingly}, we judiciously develop the prism-inspired BSP function to obtain a BSP variable $\widetilde{\mathbf{Z}}_{h}\in\mathbb{R}^{L_1 \times L_2 \times M}$, which entails the natural properties of HSIs.
{\black In other words}, the physical meaning of BSP is an unsupervised (and virtual) SSR.
Therefore, the design of our BSP ensures that the weighted sum of the two split bands is equal to the original bands, while preserving their spectral continuity. 
More details are presented in the following paragraph.
With this BSP variable, we transform the challenging SSR problem to a simpler one, i.e., the least-squares problem $\| \mathbf{Z}_h^{'}-{\widetilde{\mathbf{Z}}_{h}}\|_F^2$, followed by incorporating a deep regularization (to be defined next) to compensate some non-linear effects during the BSP.
Then, we derive the virtual HSI from the optimal solution of this deeply regularized problem [cf. \eqref{PnP vHSI}].

{\black For this purpose}, we first describe the implementation of the BSP function {\black as follows}.
Given a $P$-band MSI $\mathbf{Z}_m$, the BSP function $f_{\text{BSP}}(\cdot)$ splits each {\black spectral} band within the MSI into $\tau\in\mathbb{Z}_{++}$ bands to obtain the $M$-band ${\widetilde{\mathbf{Z}}_{h}}$ (thus, $\tau P=M$), i.e., ${\widetilde{\mathbf{Z}}_{h}}=f_{\text{BSP}}(\mathbf{Z}_m)$.
To {\black maintain} the computational efficiency while resolving the underdetermined issue, $\tau$ can be determined as 
\begin{align}\label{determine_tau}
\tau^\star:=\arg\min_{\widetilde{\tau}\in\mathbb{Z}_{++}}~\widetilde{\tau}~~\text{s.t.}~~\widetilde{\tau}P\geq N.
\end{align}
In practice, {\black representative} multispectral satellite images include the 12-band European Space Agency (ESA)'s Sentinel-2 and the 4-band Systeme Probatoire d'Observation de la Terre (SPOT) (i.e., $4\leq P\leq 12$) \cite{SPOT7}, and usually the number of materials in a multispectral remote sensing image is about in the range of $4\leq N\leq 7$ \cite{MU1}.
Thus, according to \eqref{determine_tau}, $\widetilde{\tau}:=2$ basically guarantees the overdetermined property of $\widetilde{\tau} P\geq N$.
{\black Accordingly}, we focus on the scenario of $\tau^\star:=2$, for which the two split HSI bands (from a single MSI band $[\mathbf{Z}_m]_{:,:,q}$) can always be represented as $0.5[\mathbf{Z}_m]_{:,:,q}+[\widetilde{\mathbf{H}}]_{:,:,q}$ and $0.5[\mathbf{Z}_m]_{:,:,q}-[\widetilde{\mathbf{H}}]_{:,:,q}$ for some $\widetilde{\mathbf{H}}\in\mathbb{R}^{L_1\times L_2\times P}$ (to be built below) according to the nature of light splitting.
{\black Specifically}, the $r$th HSI band can be explicitly written as
\begin{align}\label{equ:LS}
[\widetilde{\mathbf{Z}}_h]_{:,:,r}=\left\{
\begin{aligned}
&0.5\left( \left[\mathbf{Z}_m\right]_{:,:,\frac{r+1}{2}}-\left[{\mathbf{H}}\right]_{:,:,\frac{r+1}{2}} \right),&\!\!\!\!\textrm{for odd}~r,
\\
&0.5\left( \left[\mathbf{Z}_m\right]_{:,:,\frac{r}{2}}+\left[{\mathbf{H}}\right]_{:,:,\frac{r}{2}} \right),&\!\!\!\!\!\!\!\textrm{for even}~r,
\end{aligned}
\right.
\end{align}
where we {\black define} $[\mathbf{H}]_{:,:,q}:=\frac{1}{4}([\mathbf{Z}_m]_{:,:,q+1}-[\mathbf{Z}_m]_{:,:,q}),~\forall q\in\mathcal{I}_{P-1},$ for spectral continuity.
{\black For} the  boundary case $[\mathbf{H}]_{:,:,P}$, we {\black directly} compute $[\mathbf{H}]_{:,:,P}:=\frac{1}{4}([\mathbf{Z}_m]_{:,:,P}-[\mathbf{Z}_m]_{:,:,P-1}$).
Therefore, the spectral response matrix can be determined as $\bD\triangleq\bI_P\otimes\mathbf{1}^T_{\tau}$ to bridge the spectral relation of $\mathbf{Z}_m=\mathbf{Z}_h \times_3 \bD$.
{\black As a result}, if there is any negative entry in the BSP variable $\widetilde{\mathbf{Z}}_h$, it is set to zero to maintain the {\black non-negativity} of a valid HSI.
%
%%%%%%%%%%%%%%%%%%
%
Note that $\widetilde{\mathbf{Z}}_h$ is just a rough guess of the virtual HSI.
In order to make it closer to the real situation, we refine its structure using the proximal mapping, i.e.,
\begin{equation}\label{PnP vHSI}
   \mathbf{Z}_h=\arg\min_{\mathbf{Z}_h^{'}}~\| \mathbf{Z}_h^{'}-{\widetilde{\mathbf{Z}}_{h}}\|_F^2+
   \phi(\mathbf{Z}_h^{'}),
\end{equation}
where $\phi$ promotes some desired structural properties of an image, such as the sparsity or the self-similarity \cite{SSSS}.
To implement \eqref{PnP vHSI} efficiently, note that it is nothing but a denoising problem.
{\black Thus}, as the plug-and-play (PnP) strategy \cite{2024SISHY} suggests that $\phi$ can be implicit, we solve \eqref{PnP vHSI} to obtain the desired virtual HSI $\mathbf{Z}_h$, by denoising the image $\widetilde{\mathbf{Z}}_{h}$ using the fast and effective denoising CNN (DnCNN) function \cite{DnCNN}.
We notice that several DnCNN variants can be selected for our implementation.
For the sake of user-friendliness, we employ the default pretrained DnCNN network provided in MATLAB's ``Deep Learning Toolbox'' to conduct a single-step denoising for resolving \eqref{PnP vHSI}.
Moreover, since this default pretrained DnCNN is a scene-adaptive denoising network, the manual tuning for the trade-off parameter (i.e., noise level) is unnecessary in our implementation.
Thus, there is no trade-off parameter in \eqref{PnP vHSI}.  
With the above strategies, we can {\black efficiently and effectively} obtain $\mathbf{Z}_h$.
%==========================
\begin{figure}[t]
    \centering
    \includegraphics[width=1\linewidth]{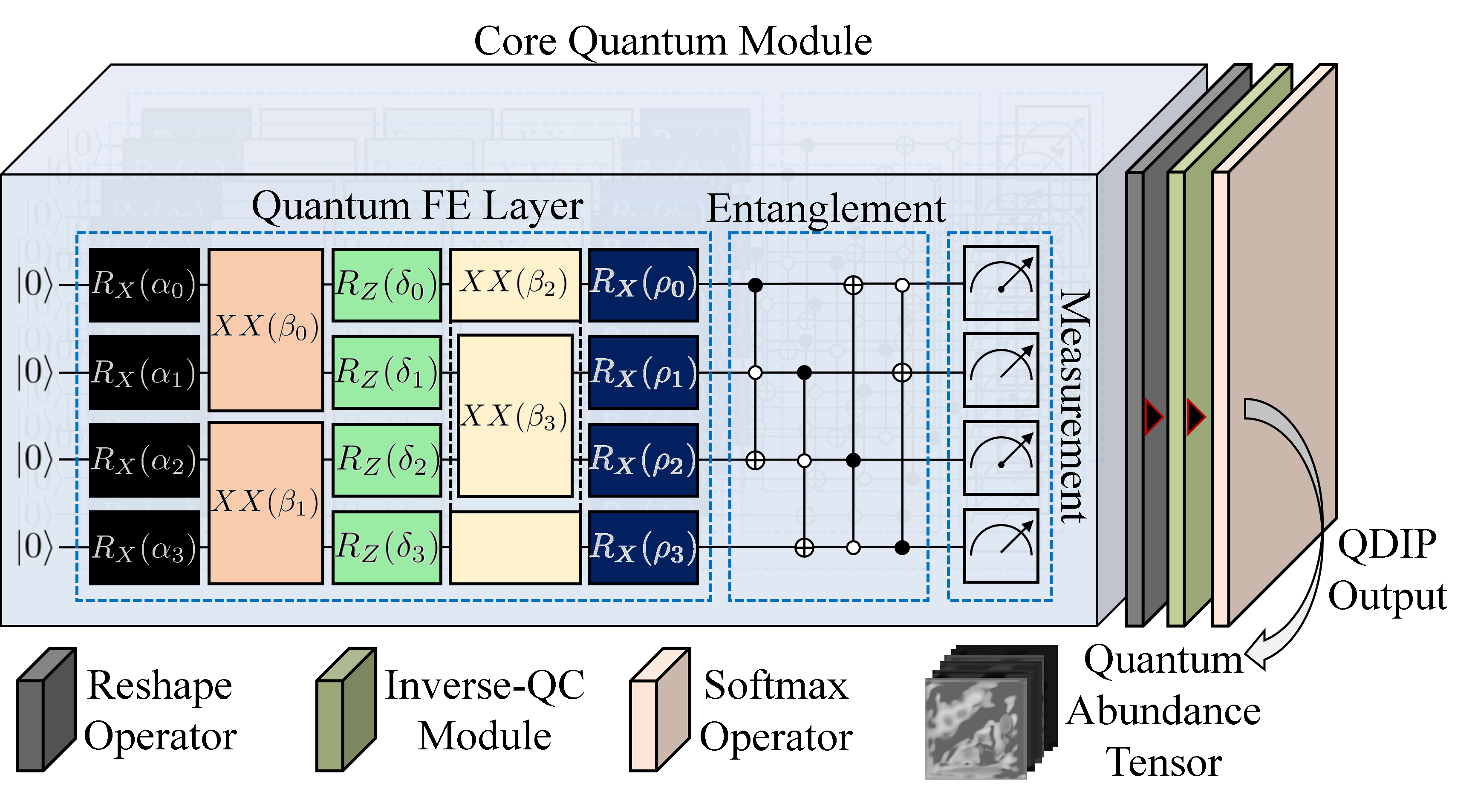}
    \caption{Graphical illustration of the proposed quantum deep image prior (QDIP), composed of a core quantum module, reshape operator, inverse quantum collapse (QC) module, and softmax operator.
    First, we use the core quantum module to obtain the quantum features.
    Specifically, the module is developed by the quantum full-expressibility (FE) layer ``$R_X-XX-R_Z-XX-R_X$'', with trainable real-valued parameters (i.e., $\alpha_i, \beta_i, \delta_i$, and $\rho_i$).
    With the provable FE, it can perform any valid quantum operators (cf. Theorem \ref{theorem: FE}).
   Also, we use the Toffoli (i.e., CCNOT) entangled layers for additional entanglement, and we employ the Pauli-Z measurement (cf. Supplementary Table I) to obtain the quantum statistics.
    Once the core quantum module holds the desired quantum states, the inverse-QC module is skillfully leveraged to prevent the valuable quantum information from being vanished during the reading-out stage.
    Eventually, the softmax operator guarantees that the output of QDIP satisfies the abundance sum-to-one property.
    More detailed configurations are summarized in Supplementary Table II.
    }
    \label{fig: Quantum Deep Image Prior}
\end{figure}
\begin{table}[t]
\centering
\caption{Quantitative evaluations of the BSS algorithms over various scenarios in terms of $\phi_{\text{en}}$ $~\!(\downarrow)$, $\phi_{\text{ab}}$ $~\!(\downarrow)$, and RMSE $~\!(\downarrow)$. We use boldfaced and underlined numbers to highlight the best and the second quantitative performance, respectively. Besides, the computational time was measured in seconds (sec.). All quantitative metrics are reported as the mean of ten independent runs; the corresponding standard deviations are presented in Supplementary Table III.}\label{tab: performance}
\begin{tabular}{cc|cccc} \hline\hline
                Methods &   Metrics      & Boulder  & Ottawa    & Hawaii     & Average        \\ \hline
\multirow{4}{*}{VCA}    & $\phi_{\text{en}}$ $~\!(\downarrow)$  & 24.442 & 17.450 & \underline{15.663} & 19.185
\\
                           & $\phi_{\text{ab}}$ $~\!(\downarrow)$    & 71.227 & 75.748 & 54.124 & 67.033  
\\
                           & RMSE $~\!(\downarrow)$  & 35.581 & 114.802 & 19.146 & 56.510  
\\
                           & Time (sec.)  & 0.009 & 0.016 & 0.009 & 0.011
\\ 
\hline
\multirow{4}{*}{NMF}   & $\phi_{\text{en}}$ $~\!(\downarrow)$  & 31.424 & 22.056 & 27.048 & 26.843  
\\
                           & $\phi_{\text{ab}}$ $~\!(\downarrow)$   & 48.784 & 44.366 & \underline{47.464} & 46.871
\\
                           & RMSE $~\!(\downarrow)$   & 28.306 & 27.223 & 25.616 & 27.048
\\
                           & Time (sec.)  & 14.463 & 14.674 & 14.463 & 14.533 
\\ 
\hline 
\multirow{4}{*}{MVC-NMF}   & $\phi_{\text{en}}$ $~\!(\downarrow)$  & 29.007 & 33.669 & 37.489 & 33.388 
\\
                           & $\phi_{\text{ab}}$ $~\!(\downarrow)$    & 53.958 & 51.863 & 62.907 & 56.243 
\\
                           & RMSE $~\!(\downarrow)$  & 26.458 & 22.790 & 28.952 & 26.067 
\\
                           & Time (sec.)   & 462.671 & 472.424 & 465.071 & 466.722 
\\ 
\hline 
\multirow{4}{*}{R-CoNMF}   & $\phi_{\text{en}}$ $~\!(\downarrow)$  & 71.392 & 63.839 & 63.460 & 66.230
\\
                           & $\phi_{\text{ab}}$ $~\!(\downarrow)$    & 50.608 & 48.348 & 53.670 & 50.875
\\
                           & RMSE $~\!(\downarrow)$  & \underline{24.847} & 21.852 & 26.597 & 24.432 
\\
                           & Time (sec.)   & 44.579 & 39.475 & 40.865 & 41.640 
\\ 
\hline 
\multirow{4}{*}{UnDIP}   & $\phi_{\text{en}}$ $~\!(\downarrow)$  & 26.364 & {\bf 3.870} & 17.701 & \underline{15.978}
\\
                           & $\phi_{\text{ab}}$ $~\!(\downarrow)$    & 55.474 & \underline{42.573} & 51.956 & 50.001 
\\
                           & RMSE $~\!(\downarrow)$  & 31.487 & \underline{13.389} & \underline{16.021} & \underline{20.299}
\\
                           & Time (sec.)  & 57.733 & 58.001 & 57.497 & 57.744 
\\ 
\hline 
\multirow{4}{*}{MiSiCNet}   & $\phi_{\text{en}}$ $~\!(\downarrow)$  & 28.693 & 11.192 & 19.481 & 19.789
\\
                           & $\phi_{\text{ab}}$ $~\!(\downarrow)$    & 60.524 & 49.966 & 65.380 & 58.623  
\\
                           & RMSE $~\!(\downarrow)$  & 28.944 & 21.848 & 31.973 & 27.588
\\
                           & Time (sec.)   & 151.434 & 140.646 & 152.659 & 148.246
\\ 
\hline 
\multirow{4}{*}{DAAN}   & $\phi_{\text{en}}$ $~\!(\downarrow)$  & \underline{19.215} & 21.446 & 16.730 & 19.130
\\
                           & $\phi_{\text{ab}}$ $~\!(\downarrow)$    & \underline{46.528} & 44.965 & 47.768 & \underline{46.420}
\\
                           & RMSE $~\!(\downarrow)$ & 25.530 & 21.591 & 22.339 & 23.153 
\\
                           & Time (sec.)   & 68.521 & 69.000 & 68.604 & 68.708 
\\ 
\hline 
\multirow{4}{*}{SSAFNet}   & $\phi_{\text{en}}$ $~\!(\downarrow)$  & 22.979 & 15.110 & 19.397 & 19.162   
\\
                           & $\phi_{\text{ab}}$ $~\!(\downarrow)$  & 59.917 & 41.666 & 68.099 & 56.561    
\\
                           & RMSE $~\!(\downarrow)$  & 32.336 & 18.962 & 34.441 & 28.580  
\\
                           & Time (sec.)   & 39.043 & 80.182 & 79.030 & 66.085 
\\ 
\hline 
\multirow{4}{*}{GQ-$\mu$}   & $\phi_{\text{en}}$ $~\!(\downarrow)$  & {\bf 11.885} & \underline{5.491} & {\bf 8.191} &  {\bf 8.522}
\\
                           & $\phi_{\text{ab}}$ $~\!(\downarrow)$    & {\bf 26.659} & {\bf 25.215} & {\bf 34.482} &  {\bf 28.785} 
\\
                           & RMSE $~\!(\downarrow)$   & {\bf 10.069} & {\bf 8.033} & {\bf 6.477} & {\bf 8.193}
\\
                           & Time (sec.)   & 51.766 & 50.194 & 50.225 & 50.728 
\\ 
\hline\hline

\end{tabular}
\end{table}
%======================================================================================

\subsection{Quantum Deep Image Prior}\label{subsec:QDIP}
This section {\black presents} the architectural design and training strategy of the proposed QDIP to extract the QAT $\mathbf{S}_{\text{qu}}$.
Before proceeding, we provide the intuition for adopting the quantum network instead of its classical counterpart (e.g., CNN) to facilitate a better {\black understanding}.
The network here (whether quantum or classical) aims to obtain a {\black approximate} solution and, hence, provide {\black an} informative prior for the proposed GQ-$\mu$ [cf. Eq. \eqref{Q-norm REG}]. 
In the underdetermined MU, the multispectral endmembers are linearly dependent (i.e., $P<N$), which may refer to the same abundance in the low-dimensional space.
For example, the pair of 1st and 6th abundances, the pair of 2nd and 3rd abundances, and the pair of 4th and 5th abundances {\black obtained by} the classical DIP (i.e., non-quantum DIP) {\black appear nearly identical}, as {\black illustrated} in Figure \ref{fig: BLDR11_result}.
Therefore, an encoder that can search for the desired abundance in the high-dimensional space potentially yields effective estimations.
{\black For intuition}, considering a typical convolutional layer with kernel size of $1\times 1$, the relation between a specific $C$-channel input noise $\bN\in\mathbb{R}^{C\times L}$ and $H$-channel output feature $\bF\in\mathbb{R}^{H\times L}$ can be generally expressed as $\bF=\sigma(\bC\bN+\bb\mathbf{1}_{L}^T)$, where $\bC^{H\times C}$, $\bb^{H\times 1}$, and $\sigma(\cdot)$ are the convolutional weights (in matrix-multiplication form), bias vector, and the activation function, respectively.
Typically, the default setting of $\bb$ is an all-zeros vector, and the ReLU activation function is adopted in DIP. 
When a convolutional network is composed of $K$ layers, it hardly guarantees that all convolutional kernel matrices remain full-rank during the training procedure, and may result in rank deficiency in classical convolutional networks.
While the above analysis just shows an approximation for knowing the intuition, similar conclusions of the restrictions in classical networks {\black have} also been reported in the literature \cite{ansuini2019intrinsic,recanatesi2019dimensionality,feng2022rank}.
This explains why classical DIP hardly transfers the problem into an overdetermined one.

Conversely, quantum machine learning (QML) provides a natural advantage in mapping classical data into a higher-dimensional space (e.g., Hilbert space) \cite{caro2022generalization}.
QML also excels at capturing the complex feature interactions to facilitate more abstract feature {\black representations} via quantum entanglement (which is also the key behind quantum small-data machine learning) \cite{wang2024quantum}.
{\black By} integrating these advantages {\black with} the provable full expressibility (FE, as shown in Theorem \ref{theorem: FE}) of our quantum network, the proposed quantum FE module enables searching for the target state throughout the higher-dimensional superposition quantum space \cite{wang2024quantum} under a single-data setting.
The architecture of the QDIP is {\black illustrated} in Figure \ref{fig: Quantum Deep Image Prior}, where the detailed configuration is summarized in Supplementary Table II due to {\black space limitations}.
First, the QDIP {\black consists of} the core quantum module, which is developed with provable FE, followed by the entanglement and measurement layers.
The mathematically guaranteed FE ensures that the quantum FE layer can realize any valid quantum unitary operators (cf. Theorem \ref{theorem: FE}); note that all the valid quantum operators must be unitary.
Accordingly, we believe that the core quantum module can {\black represent} any desired quantum image state $\mathbf{S}_{\text{qu}}$; this is why the proposed QDIP needs no specific input, just like the conventional DIP \cite{DIP2018}.
Once the core quantum module holds the desired quantum state, we use the inverse quantum collapse (inverse-QC) module \cite{HyperQUEEN} composed of transposed convolution (TConv) blocks (cf. Supplementary Table II) to {\black preserve} the valuable quantum information from being collapsed during the quantum read-out procedure (cf. Figure \ref{fig: Quantum Deep Image Prior}).
Finally, the softmax operator guarantees the output QAT to {\black preserve} the abundance sum-to-one nature.
{\black For completeness}, the overall configurations of the QDIP are detailed in Supplementary Table II.

Subsequently, we introduce the design of the core quantum module.
We summarize the {\black employed} quantum gates {\black along with} their corresponding unitary operators in Supplementary Table I.
As {\black illustrated} in Figure \ref{fig: Quantum Deep Image Prior}, the core quantum module is {\black constructed based on} the FE architecture ``$R_X-XX-R_Z-XX-R_X$'', i.e., the Rotation-Ising quantum FE layer.
To {\black exploit} the quantum entanglement attribute, we also employ the Toffoli (i.e., CCNOT) entangled layers for additional entanglement, followed by the Pauli-Z quantum measurement.
{\black Notably}, this light-weight design of the core quantum module is well {\black aligned with} the near-term quantum computers, because even some of the most advanced quantum computers (e.g., the 127-qubit IBM Eagle) have very limited qubit resources \cite{Eagle}.
The core quantum module, even containing just a hundred-scale qubits, can express any valid quantum unitary function:
\begin{theorem}\label{theorem: FE} 
The trainable quantum neurons deployed in the quantum FE layer embedded within the core quantum module (cf. Figure \ref{fig: Quantum Deep Image Prior}) can express any valid quantum unitary operator $U$, with some real-valued trainable network parameters $\{\alpha_i, \beta_i, \delta_i, \rho_i \}$. 
\hfill$\square$
\end{theorem}
The proof of Theorem \ref{theorem: FE} is {\black provided} in Supplementary Note 1 due to space limitations.
Instead, we {\black briefly discuss} the key concept of quantum expressibility.
For a specific parameterized quantum circuit (PQC), such as QUEENs, its expressibility is {\black characterized} by the ability to explore the target quantum states in the quantum space \cite[Section 3.1]{Expressibility}. 
In the single-qubit PQC {\black case}, the expressibility is the capability for exploring the Bloch sphere.
A typical quantitative assessment of the expressibility is the comparison between the distribution of sampling states and the uniform distribution of states \cite{Expressibility}.
A higher expressibility may not necessarily yield a better performance, while restricted expressibility indeed results in limited performance.
However, investigating the expressibility of all kinds of PQCs is impractical during architecture development.
Therefore, we develop the QUEEN with the theoretically guaranteed full expressibility (FE).
This design ensures {\black that} our QUEEN can realize any valid quantum operators for the challenging abundance estimation.

Next, {\black we} discuss the training of QDIP.
We explicitly express the output of QDIP as $\widetilde{\mathbf{S}}_{\text{qu}}:=\text{Q}_{\theta}(\cdot)$, where $\text{Q}_{\theta}(\cdot)$ denotes the overall function of QDIP with $\theta$ being the network parameters.
As discussed {\black earlier}, {\black we} can obtain the desired QAT $\mathbf{S}_{\text{qu}}$ using the proposed QDIP, i.e.,
\begin{equation}\label{subproblem QNN}
\theta^{\star}\in \arg\min_{\theta}~\mathcal{L}_{\text{QDIP}}(\widetilde{\mathbf{S}}_{\text{qu}}); ~~~\mathbf{S}_{\text{qu}}:=\text{Q}_{\theta^{\star}}(\cdot),
\end{equation}
where $\mathcal{L}_{\text{QDIP}}$ and $\theta^{\star}$ denote the task-specific loss function and the optimal network parameters of QDIP, respectively.
Inspired by \cite{UnDIP} and LMM, {\black we design} the loss function as
\begin{equation}\label{loss1 QNN} 
\mathcal{L}_{\text{QDIP}}(\widetilde{\mathbf{S}}_{\text{qu}})=\|\mathbf{Z}_h-\widetilde{\mathbf{S}}_{\text{qu}} \times_3 \bA^{0}\|_F^2+\text{QREG}(\widetilde{\mathbf{S}}_{\text{qu}}),
\end{equation}
where the {\black regularization term} $\text{QREG}(\cdot)$ is {\black implicitly induced by} QDIP network itself (cf. Figure \ref{fig: Quantum Deep Image Prior}) \cite{DIP2018}, {\black while} $\mathbf{Z}_h$ and $\bA^{0}$ are {\black specified} in Section \ref{subsec:Virtual HSI} and Section \ref{subsec:Closed-form}, respectively.
The quantum neurons are unitary operators (cf. Supplementary Table I) instead of affine mappings (e.g., convolution), so QDIP provides unique quantum features for abundance estimation [cf. \eqref{Q-norm REG}].
We employ the adaptive moment estimation (Adam) \cite{Adam} to optimize \eqref{subproblem QNN}-\eqref{loss1 QNN}, with detailed training settings summarized in Section \ref{subsec:Experimental Protocol}.

For the first time, the quantum version of the seminal idea of DIP \cite{DIP2018} is proposed above.
For a more comprehensive introduction to quantum deep learning, including the Dirac notation system, barren plateaus (BP) effect (a.k.a. quantum gradient vanishing) and quantum measurement/collapse, we refer readers to \cite[Section II.A]{HyperQUEEN} and \cite[Section II.B]{HyperQUEEN}.

\section{Experimental Evaluation}\label{sec:Experiments}
In this section, we conduct {\black extensive} experiments to evaluate the performance of the proposed GQ-$\mu$.
{\black For this purpose}, we design an experimental protocol (cf. Section \ref{subsec:Experimental Protocol}) for the underdetermined MU task {\black due to} the lack of reference datasets.
{\black Moreover}, the experimental settings (including the proposed GQ-$\mu$ and baselines) are also {\black described} in Section \ref{subsec:Experimental Protocol}.
Next, the qualitative and quantitative evaluations are {\black presented} in Section \ref{subsec:Evaluations}.
Section \ref{subsec:Ablation Study} {\black reports} ablation studies to substantiate the effectiveness of the proposed QDIP [cf. \eqref{Q-norm REG}] and WSS regularization [cf. \eqref{WSS}].
Section \ref{subsec: Real_data} presents the real-data analysis to demonstrate the {\black practical applicability} of our GQ-$\mu$.
Besides, parameter and convergence analysis are provided in Supplementary Note 2 due to {\black space limitations}.
\subsection{Experimental Protocol Design and Settings}\label{subsec:Experimental Protocol}
\begin{figure}[t]
    \centering
    \includegraphics[width=0.98\linewidth]{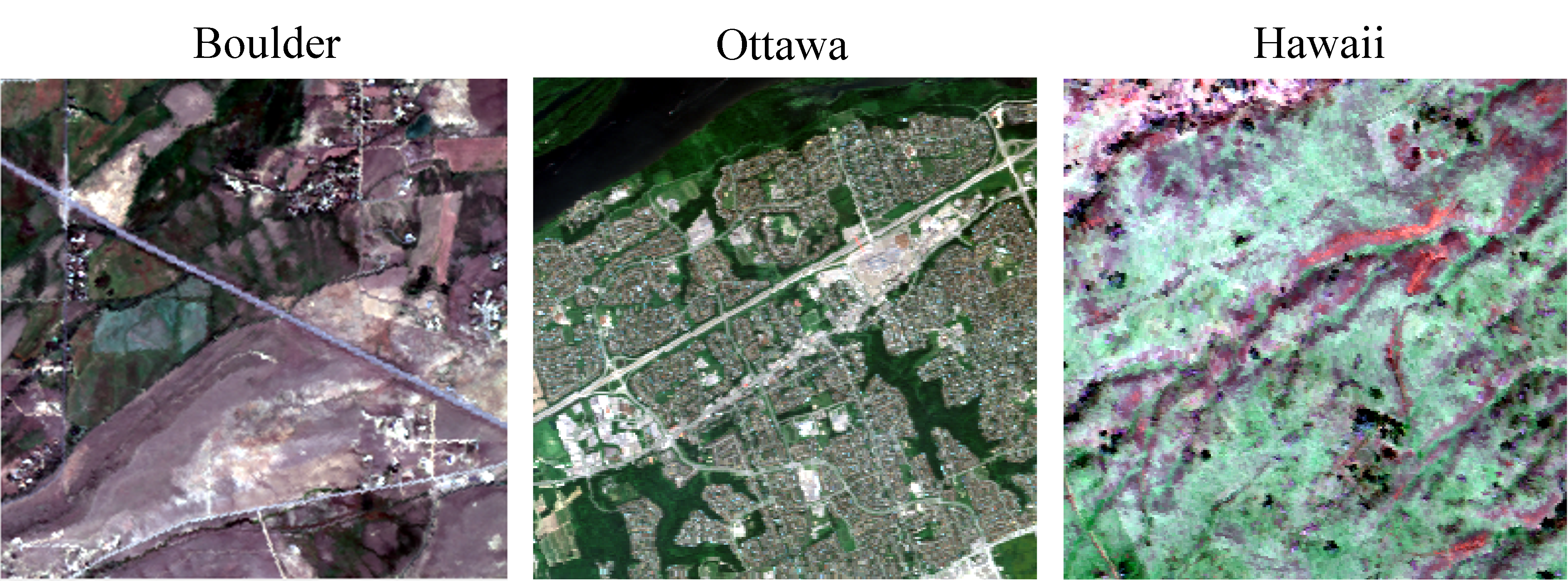}
    \caption{False-color compositions (bands 25, 12, and 6 as RGB) of NASA's HSIs captured by the AVIRIS sensor for different locations and landscapes.
    }
    \label{fig: ROI}
\end{figure}
\begin{figure*}[t]
    \centering
    \includegraphics[width=0.97\linewidth]{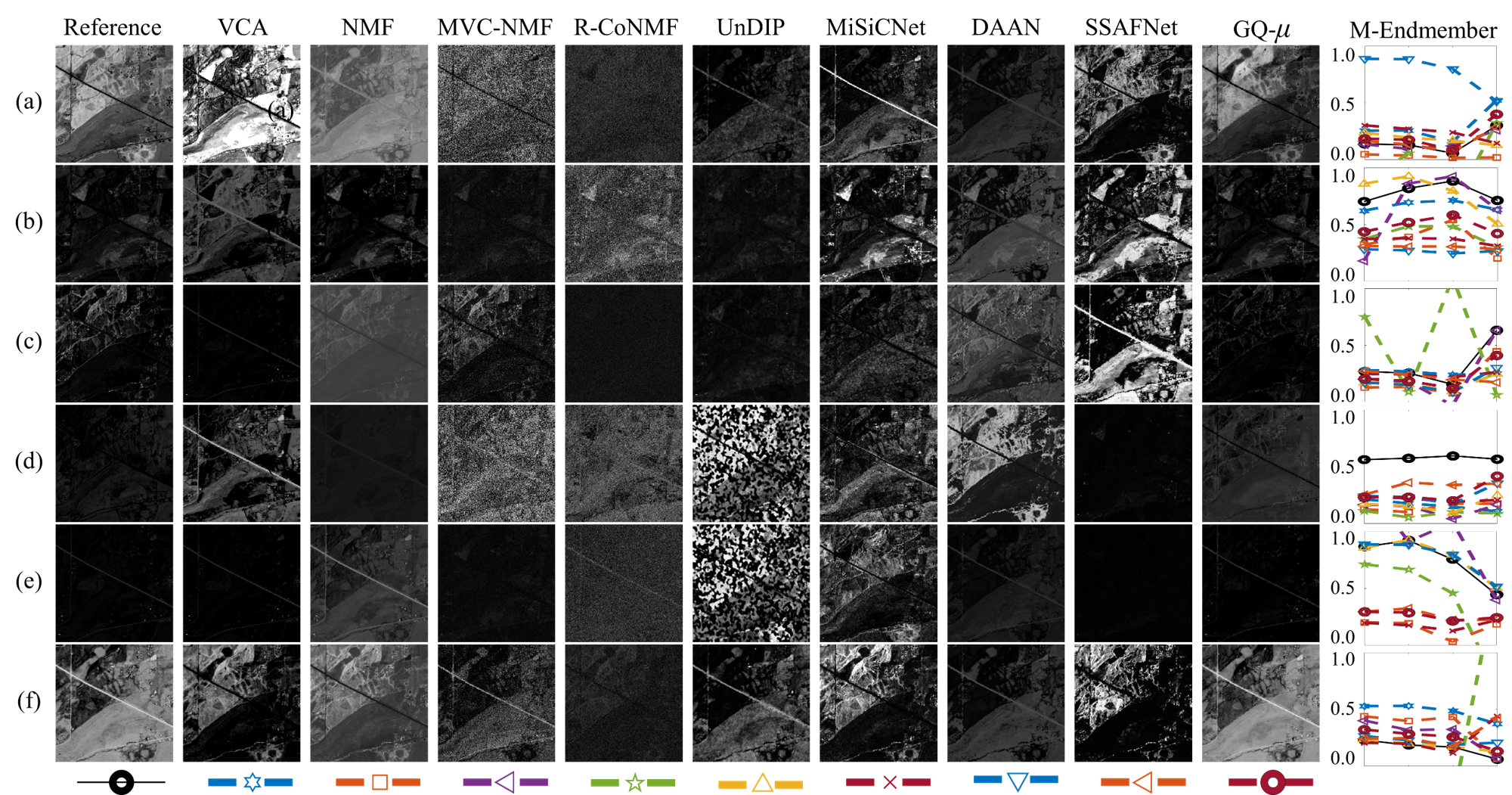}
    \caption{The qualitative comparisons over the Boulder dataset. The corresponding multispectral endmembers (i.e., M-endmember) of the $n$th abundance are shown in the $n$th subfigure within the {\black rightmost column, where the horizontal and vertical axes represent the spectral bands and reflectance values, respectively.}  
    }
    \label{fig: BLDR11_result}
\end{figure*}
\begin{figure*}[t]
    \centering
    \includegraphics[width=0.97\linewidth]{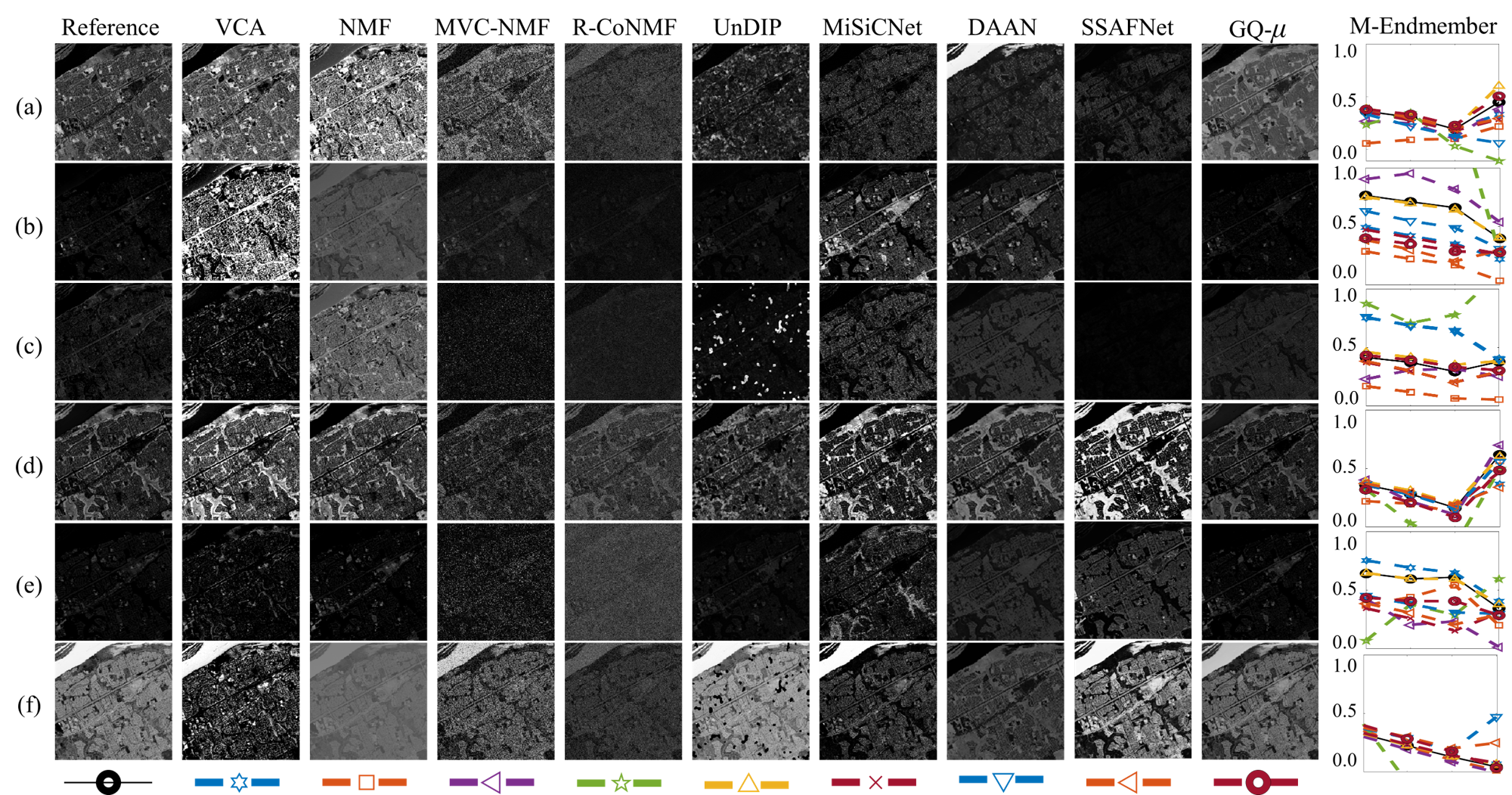}
    \caption{The qualitative comparisons over the Ottawa dataset. The corresponding multispectral endmembers (i.e., M-endmember) of the $n$th abundance are shown in the $n$th subfigure within the {\black rightmost column, where the horizontal and vertical axes represent the spectral bands and reflectance values, respectively.}  
    }
    \label{fig: ottawa_result}
\end{figure*}
Initially, we {\black observe} that the previous MU studies do not consider the underdetermined issue; thus, experimental datasets or protocols for MU are unavailable.
{\black For systematic evaluation}, we design the experimental protocol for this {\black challenging} MU task, which is {\black illustrated} in Supplementary Figure 1 due to space limitations.
Specifically, inspired by Wald’s protocol for hyperspectral restoration or image fusion \cite{COCNMF}, we {\black adopt} a similar approach to obtain the reference endmember matrix $\bB$ and abundance tensor $\mathbf{S}$.
In this protocol (cf. Supplementary Figure 1), we first perform {\black an} effective BSS algorithm (e.g., HISUN \cite{HISUN}) on the reference HSI, thereby obtaining the reference hyperspectral signature $\bA_{\text{ref}}$ and the reference abundance tensor $\mathbf{S}$.
{\black Subsequently}, we uniformly downsample the reference hyperspectral signature $\bA_{\text{ref}}$ along the spectral dimension (to be described next) to obtain the reference endmember matrix $\bB$ \cite{PRIME}.
{\black Accordingly}, we synthesize the reference MSI, i.e., $\mathbf{Z}_{m}=\mathbf{S}\times_3\bB+e\mathbf{N}_{\text{de}}$, where $\mathbf{N}_{\text{de}}\in\mathbb{R}^{L_1 \times L_2 \times P}$ is the normally distributed LMM deviation tensor; and $e:=$1E-4 is the scaling factor determined by the energy of the LMM deviation of reference HSI.
Once we have $\mathbf{Z}_{m}$, the estimated signature $\widehat{\bB}$ and abundance $\widehat{\mathbf{S}}$ can be obtained by Algorithm \ref{alg:MUmain} or other benchmarks.
Eventually, the qualitative and quantitative evaluations between $\bB$ and $\widehat{\bB}$ (resp. $\mathbf{S}$ and $\widehat{\mathbf{S}}$) can be {\black readily conducted} (cf. Supplementary Figure 1).
{\black Moreover}, we also experimentally replace $\mathbf{N}_{\text{de}}$ with a residual-based noise tensor.
To approximate realistic sensor noise characteristics, we first {\black perform} spectrally {\black downsampling} on the reference HSI (using the same approach as downsampling $\bA_{\text{ref}}$) to approximate the noisy MSI.
Then, we apply the pretrained DnCNN denoiser on the noisy MSI and {\black treat} the residual as a real-world noise tensor.
With the quantitative metrics (cf. Section \ref{subsec:Evaluations}) and default {\black parameter} settings (to be detailed later), our GQ-$\mu$ achieves an average quantitative performance of $\phi_{\text{en}}=8.249$, $\phi_{\text{ab}}=28.768$, and $\text{RMSE}=8.166$ with normally distributed LMM deviation tensor.
{\black In contrast}, it achieves $\phi_{\text{en}}=7.266$, $\phi_{\text{ab}}=28.080$, and $\text{RMSE}=7.385$ with residual-based noise tensor.

To {\black implement} the designed protocol, we adopt the HSIs captured by NASA’s Airborne Visible/Infrared Imaging Spectrometer (AVIRIS) sensor \cite{AVIRISdata}. 
These HSIs are originally composed of 224 contiguous bands, while we remove those water-vapor absorption bands (i.e., bands 1-10, 104-116, 152-170, and 215-224), followed by using the remaining 172 clean bands as the reference HSIs for the experimental protocol (cf. Supplementary Figure 1). 
For the spectral downsampling stage of $\bA_{\text{ref}}$, we select the spectral ranges covering 450–520, 520–600, 630–690, and 760–900 nm for synthesizing the MSIs.
{\black These wavelength ranges correspond to the Landsat TM satellite bands 1–4 and are consistent with the red-green-blue (RGB) and near-infrared (NIR) bands provided by typical multispectral imagery systems.
The NIR band exhibits distinct spectral characteristics compared to RGB \cite[Table 3]{10530973}, thereby enabling the material identification of multispectral data to some extent.}
Furthermore, according to the spectral continuity, the reflectances in these hyperspectral bands are averaged (for each band range) to approximate the multispectral reflectance (of each band) instead of directly choosing/quantizing a specific band.
These approximated multispectral reflectances enable us to consider the spectral reflectance of the objects, thereby obtaining the realistic reference endmember matrix $\bB$ ($P=4$).
Remarkably, even if we have very few MSI bands, the literature often used the ``multispectral shape'' to distinguish objects in, for example, precision agriculture.
That said, for MSI, we still have the concept of reflectance endmember signatures \cite{MU1}.
{\black In this experiment, we empirically set $N:=6$, and the comparisons of the multispectral endmembers within a single ROI are presented in Supplementary Figure 2.
Despite the limited number of bands, the multispectral endmembers exhibit distinct reflectance with non-overlapping shapes, indicating their material identifiability.
However, when multispectral observations are insufficient for a more meticulous spectral analysis, converting them to their corresponding hyperspectral counterparts (cf. Supplementary Figure 3) via customized algorithms \cite{COS2A} can provide improved practicality.
Subsequently, the descriptions of the used HSIs for data simulation are summarized below.}
For {\black extensive} evaluations, we select three HSIs captured over different locations (cf. Figure \ref{fig: ROI}); and all {\black datasets} have $L=256\times256$ pixels.
Specifically, the first dataset (i.e., Boulder) was acquired over Boulder, Colorado, with a spatial resolution (SR) of 10.7 m.
The second dataset (i.e., Ottawa) was captured over Ottawa, Canada, with a SR of 17.1 m; and the third dataset (i.e., Hawaii) was captured over Hawaii County, Hawaii, with a SR of 2.2 m.
Next, we {\black describe} the benchmarks and their parametric settings.
{\black Notably}, this is a pioneering study for underdetermined MU; thus, there is no existing algorithm for this challenging task.
{\black Accordingly}, we extend several representative HU algorithms to compare with the proposed GQ-$\mu$, and the details are {\black described below}.
{\black First}, we {\black adapt} the representative approaches, i.e., VCA \cite{VCA} and NMF \cite{NMF-peer}, to compare with our approach \cite{MU1}.
Specifically, VCA originally performs the dimensional reduction (DR) for HU, while we {\black directly} conduct the algorithm in the original multispectral domain because of the underdetermined issue (i.e., $P<N$).
{\black As} VCA only estimates the signature $\widehat{\bB}$, the abundance of VCA {\black is obtained} by left-multiplying the pseudo-inverse of the estimated signature $\widehat{\bB}^\dag$ on the MSI (i.e., $\widehat{\mathbf{S}}=\mathbf{Z}_m \times_3 \widehat{\bB}^\dag$), {\black as suggested in} the original VCA paper.
In NMF, we {\black adopt} HISUN \cite{HISUN} as a warm start strategy to initialize $\bB$ and $\mathbf{S}$; and we set the number of iterations to 1E+3 {\black to improve} convergence.
{\black After} NMF converges, we refine the abundance by solving the non-negative least-squares optimization problem of the returned signature and the MSI, which yields an {\black improved} abundance than the returned one.
Moreover, we {\black additionally} incorporate two {\black well-known} simplex-volume regularized NMF approaches, i.e., minimum volume constrained NMF (MVC-NMF) \cite{MVCNMF} and robust collaborative NMF (R-CoNMF) \cite{CoNMF}, with spectral augmentation to conduct the underdetermined BSS.
In detail, these approaches are also designed for HU and, hence, inevitably require the DR basis matrix for the gradient descent procedure during the alternative optimization (AO).
To address the underdetermined issue (i.e., $P<N$), we first apply the BSP function {\black defined in} \eqref{equ:LS}, followed by adding a small amount of normally distributed perturbation to resolve the rank deficiency.
This yields the full-rank $M$-band (with $M>N$) input for these methods (i.e., MVC-NMF and R-CoNMF).
After estimating $M$-band signatures $\widehat{\bA}$ (also the corresponding abundance $\widehat{\bS}$) using these methods, we consequently obtain the estimated multispectral endmembers by {\black performing} the spectral downsampling, i.e., $\widehat{\bB}=\bD\widehat{\bA}$.
For the parameter settings, we set the trade-off parameters $\tau:=\text{1E-3}$, tolerance factor of 1E-6, and the maximum iteration number of 5E+2 for MVC-NMF.
In R-CoNMF, we set the $\alpha:=\text{1E-6}\times \sqrt{(N/1000)}$, $\beta:=\text{5E-1}\times(N/1000)$, $\lambda_t:=(N/1000)$, $\mu_t:=1$, $\theta:=N\times\text{1E-3}$, and the number of AO to 2E+2, which are all the default settings from the official implementation.
Beyond the {\black conventional} approaches, we also employ numerous representative and state-of-the-art deep learning approaches, including HU using DIP (UnDIP) \cite{UnDIP}, minimum simplex convolutional network for deep HU (MiSiCNet) \cite{MiSiCNet}, deep autoencoder-based augmented network (DAAN) \cite{DAAN}, and spatial-spectral adaptive fusion network (SSAFNet) \cite{SSAFNet}; the corresponding experimental settings are {\black described} below.
In UnDIP, we defaultly set the number of iterations to 3E+3; and we set the learning rate (LR) to 1E-5 for training stability.
In MiSiCNet, we set the number of iterations, LR, and the trade-off parameter to 8E+3, 1E-3, and 1E+2, respectively, across all datasets.
In DAAN, the LRs of the model encoder and decoder are set to 1E-4 and 1E-2, respectively, under 8E+2 training iterations.
In SSAFNet, we set a small LR of 1E-4 to prevent the gradient vanishing, while the trade-off parameters of the loss function are defaultly set to $\lambda_{\text{KL}}:=0.001$, $\lambda_{\text{SAD}}:=5$, and $\lambda_{\text{vol}}:=10$.
{\black For} the proposed GQ-$\mu$, we first set the LR and the number of iterations of the QDIP (cf. Section \ref{subsec:QDIP}) to 5E-2 and 2E+2, respectively, to extract $\mathbf{S}_{\text{qu}}$; and the parametric settings of Algorithm \ref{alg:MUmain} are {\black summarized} below.
First, one can notice that the number of pixels is much larger than the number of sources (i.e., $L\gg N$), and hence we define $\lambda_3^{\dag}\triangleq \frac{\lambda_3}{\rm 1E4}$ and $\lambda_4^{\dag}\triangleq \frac{\lambda_4}{\rm 1E4}$.
Accordingly, we set $\lambda_1:=$1E-3, $\lambda_2:=$1E-2, and $\lambda_3^{\dag}:=$1E-1. 
As for $\lambda_4^{\dag}$, we first initialize it to 1E-2, followed by increasing $\lambda_4^{\dag}$ by 1.2 times in each iteration for a stable optimization.
Besides, we set the ADMM penalty parameter $\mu:=1$ for Algorithm \ref{alg:eq:subproblem ADMM}.
Finally, we set the number of iterations to 5 for GQ-$\mu$; and set it to 20 for Algorithm \ref{alg:eq:subproblem ADMM}.

All benchmarks are conducted on the computational equipment {\black equipped} with an NVIDIA RTX 3090 GPU, AMD Ryzen 5950X CPU {\black operating at} 3.40-GHz, 24-GB RAM, as well as with Core-i7-11700 CPU and 128 GB RAM.
The software environments for this experiment are MATLAB R2023a, Python 3.10.9, and Torch 2.0.0.
{\black For} the proposed approach, we train the QDIP on the same equipment as benchmarks, {\black and subsequently} perform the Algorithm \ref{alg:MUmain} on a laptop with a Core-i7-11700 CPU 2.50-GHz and speed 24-GB RAM.

\subsection{Qualitative and Quantitative Evaluations}\label{subsec:Evaluations}
This section {\black presents} comprehensive qualitative and quantitative evaluations.
First, we graphically {\black illustrate} the reference abundances ${\mathbf{S}}$ (resp., estimated abundances $\widehat{\mathbf{S}}$) and the reference sources $\bB$ (resp., estimated sources $\widehat{\bB}$) for Boulder, Ottawa, and Hawaii datasets in Figure \ref{fig: BLDR11_result}, Figure \ref{fig: ottawa_result}, and {\black Supplementary Figure 4 (due to space limitation)}, respectively.
In these figures, the $n$th row presents the qualitative comparisons for the $n$th abundance and the associated $P$-band sources, where the color legends are shown at the bottom of each abundance column.
With these in mind, the qualitative evaluations are presented as follows.
As mentioned {\black earlier}, we pay more attention to the abundances than the {\black endmember} signatures in typical MU.
{\black Consequently}, we first conduct qualitative analysis for the abundance over the Boulder data, {\black with} results illustrated in Figure \ref{fig: BLDR11_result}.
As shown in Figure \ref{fig: BLDR11_result}, the proposed GQ-$\mu$ yields the most precise abundance (cf. the last abundance column of Figure \ref{fig: BLDR11_result}) {\black among all} benchmarks.
Conversely, the benchmarks {\black exhibit} unsatisfactory {\black performance} for the abundance in the Boulder data, {\black characterized by} indistinguishable abundances for distinct sources, severe noise-like patterns, and unexpected structural patterns. 
For {\black example}, this indistinguishability can be {\black clearly} observed in the 3rd and 5th abundances of VCA, the 2nd and 3rd abundances of UnDIP, the 1st and 5th abundances of DAAN, and the 1st and 6th abundances of SSAFNet.
We further {\black note} that the qualitative performance of the simplex volume-based NMF approaches is {\black severely} affected by noise-like patterns; see, e.g., the 1st and 4th abundances of MVC-NMF, and the 1st, 2nd, and 4th abundances of R-CoNMF.
This may be attributed to {\black the fact} that, even though the spectral augmentation addresses the rank deficiency issue, the theoretical formulations of MVC-NMF and R-CoNMF primarily consider the hyperspectral structure and {\black fail to exploit} the spatial coherence of its multispectral counterpart. 
In addition, we note that the abundances of UnDIP have unexpected patterns (e.g., the 4th and 5th abundance of UnDIP), {\black indicating} the instability and unreliability of the ``black-box'' BSS model. 
Beyond abundance evaluations, we further conduct qualitative comparisons for the estimated multispectral signatures.
As shown in the last column of Figure \ref{fig: BLDR11_result}, the proposed GQ-$\mu$ has numerous signatures well-matched with reference signatures; e.g., see the 1st, 2nd, 3rd, and 6th signatures.
In comparison, the signatures estimated from the benchmarks achieve severe shape and intensity discrepancies; e.g., see the 3rd and 6th signatures of R-CoNMF and the 1st signature of DAAN.
From the above evaluations, both representative NMF and state-of-the-art deep learning HU approaches struggle with MU in the underdetermined scenarios. 
{\black In contrast}, the proposed GQ-$\mu$ can effectively tackle the underdetermined MU in its hyperspectral counterpart, i.e., HU (cf. Section \ref{subsec:Criterion Design}).
Furthermore, the proposed WSS and quantum regularizers effectively address the ill-posedness of HU.
Consequently, the proposed GQ-$\mu$ outperforms the baselines on the Boulder dataset.

Next, we further {\black validate} the effectiveness of our approach on the Ottawa dataset, where the qualitative results are {\black presented} in Figure \ref{fig: ottawa_result}.
The results indicate that both traditional and deep learning approaches remain limited {\black in terms of} performance (e.g., abundance indistinguishability) due to the underdetermined {\black setting}.
For example, the 1st, 2nd, and 4th abundance of VCA, and the 2nd, 3rd, and 6th abundance of NMF, the 1st and 6th abundance of R-CoNMF, and the 2nd and 3rd abundance of SSAFNet.
Although the baseline methods can partially {\black yield} relatively {\black accurate} estimations for abundance [see Figure \ref{fig: ottawa_result}(d)], the overall performance remains {\black limited}.
{\black Notably}, the undesirable patterns are again observed in the 3rd abundance of UnDIP.
Furthermore, the sum-to-one constraint within UnDIP may further propagate this distortion {\black effect} from the 3rd abundance to the 6th abundance.
In contrast, the proposed GQ-$\mu$ yields substantially promising estimated abundances as shown in the last abundance column of Figure \ref{fig: ottawa_result}, which {\black closely match} the reference abundances.
Additionally, the proposed GQ-$\mu$ provides numerous precise estimations for signatures on the Ottawa datasets; e.g., see the 1st, 3rd, 4th, and 6th signatures.
We acknowledge that some of the signatures {\black estimated by} GQ-$\mu$ may occasionally exhibit obvious imperfections; however, severe shape and intensity deviations {\black rarely occur}, such as the 3rd signature of VCA and R-CoNMF.  
In summary, our GQ-$\mu$ yields the best performance compared to the baselines on the Ottawa dataset.

{\black Finally}, we perform the qualitative comparisons for the Hawaii dataset, where the estimated abundance and signatures are illustrated in {\black Supplementary Figure 4 due to space limitations}.
{\black As expected}, the proposed GQ-$\mu$ yields promising qualitative performance in abundance estimations.
Nearly all estimated abundances from the proposed GQ-$\mu$ are closely matched to the reference abundance.  
Conversely, the traditional benchmarks (i.e., VCA and NMF) still struggle with the underdetermined issue and fail in the abundance estimation.
For example, the 1st, 3rd, and 5th abundance of VCA, and the 2nd, 3rd, 4th, and 5th abundance of NMF.
{\black Moreover}, the noise-like patterns are observed in the estimated abundances from the simplex-volume regularized NMF approaches; e.g., see the 1st, 2nd, and 5th abundance of MVC-NMF and R-CoNMF.
Beyond these approaches, the deep learning methods suffer from the underdetermined issue and result in similar abundances for different sources.
For instance, the 3rd and 5th abundance of UnDIP, the 2nd, 3rd, 5th abundance of DAAN, and the 1st, 2nd, 3rd, and 5th abundance of SSAFNet. 
According to the above evaluations, the results {\black strongly} support that our GQ-$\mu$ achieves the best qualitative performance on all the datasets.
%
%

% %
Furthermore, we {\black conduct} quantitative evaluations to fairly evaluate the results of MU.
{\black For this purpose}, we first employ a representative metric for endmembers {\black evaluation}, i.e., root-mean-square (RMS) spectral angle mapper for endmember $\phi_{\text{en}}$, whose definition {\black is given} in \cite[Eq. (32)]{HyperCSI}.
This metric quantitatively measures the RMS angle between the reference signatures and estimated signatures.
We {\black further} adopt two representative metrics for {\black abundance evaluation}, i.e., RMS error (RMSE) and RMS spectral angle mapper for abundance $\phi_{\text{ab}}$, whose definitions {\black are provided} in \cite[Eq. (14)]{UnDIP} and \cite[Eq. (33)]{HyperCSI}, respectively.
The RMSE and $\phi_{\text{ab}}$ quantitatively measure the Euclidean and cosine distance, respectively, between the reference and estimated abundance.
To {\black mitigate} the randomness of each BSS approach from biasing the quantitative evaluation, the quantitative metrics summarized in Table \ref{tab: performance} are reported as the {\black average over} ten independent runs.
Furthermore, the standard deviations of the quantitative metrics are provided in the Supplementary Table III, and the quantitative evaluations are as {\black summarized} follows.

According to Table \ref{tab: performance}, the proposed GQ-$\mu$ achieves the best quantitative performance in both abundance and signatures on the Boulder and Hawaii datasets, aligning with the qualitative evaluations.
Furthermore, we {\black note} that the quantitative performance of the runner-up methods (i.e., UnDIP and DAAN) is {\black substantially} inferior to our approach (cf. Table \ref{tab: performance}).
This observation indicates that, even with the state-of-the-art deep learning techniques, MU can not be effectively addressed without considering its ``underdetermined'' nature.
In contrast, we effectively tackle the underdetermined MU in its hyperspectral counterpart, i.e., HU.
Subsequently, the ill-posedness of HU is effectively resolved using novel quantum and WSS regularizers.
In the proposed GQ-$\mu$, the WSS regularizer considers both ``direction'' and ''force'' to recover the minimum volume data-enclosing simplex, instead of simply minimizing the data volume, such as MVC-NMF and R-CoNMF.
Additionally, the quantum regularizer effectively extracts the informative structure to estimate the true abundance rather than using merely the naive sparse prior (e.g., $\ell$-1 norm regularization). 
These collectively yield the superior performance (cf. Table \ref{tab: performance}) on the challenging MU applications.
On the Ottawa dataset, the quantitative results show that UnDIP is slightly better than our approach in terms of $\phi_{\text{en}}$, whereas its abundance estimation is unsatisfactory in terms of $\phi_{\text{ab}}$ and RMSE, consistent with the qualitative evaluations.
By comparison, the proposed GQ-$\mu$ achieves the best performance in terms of $\phi_{\text{ab}}$ and RMSE on the Ottawa data (cf. Table \ref{tab: performance}).
Hence, the proposed GQ-$\mu$ still yields the best average quantitative performance compared to the representative baselines.
As a pioneering study of underdetermined MU, we have {\black demonstrated} the feasibility of this task based on the above comprehensive experiments.
In the next section, we {\black present} an ablation study to evaluate the effectiveness of the proposed quantum and WSS regularizers.
\subsection{Ablation Study}\label{subsec:Ablation Study} 
In this section, we {\black conduct} the ablation studies to examine the effectiveness of the quantum and WSS regularizers (REGs). 
First, we {\black replace} the proposed WSS REG [see \eqref{WSS}] with various types of minimum volume (MV) REGs.
{\black Specifically}, given the hyperspectral endmember matrix $\bA=[\ba_1,\dots,\ba_N]$, the simplex volume representation {\black can be} expressed as $\text{vol}(\text{con}(\bA))=\frac{1}{(N-1)!} \left( \det(\bar{A}^{T}\bar{A}) \right)^{1/2}$ with $\bar{\mathbf{A}} \triangleq 
\left[ \mathbf{a}_1 - \mathbf{a}_N, \; \ldots, \; \mathbf{a}_{N-1} - \mathbf{a}_N \right]$ \cite{COCNMF}. 
Considering the {\black computational complexity} of the determinant and potential gradient descent procedure \cite{MVCNMF}, we {\black instead} adopt its convex surrogate, sum-of-squared distances (SSD) REG \cite[Eq. (6)]{COCNMF}, as the first baseline MV REG.
We {\black additionally} incorporate three additional quadratic baseline MV REGs referring to ``{Boundary}'', ``{Center}'', and ``{TV}'' into this comparison.
The explicit {\black formulations} of these quadratic MV REGs can be found in \cite[Table I]{centerMVES}.
Finally, by setting all the weights to one (i.e., $\bw:=\mathbf{1}_N$), the proposed WSS REG is simplified to its non-weighted counterpart to evaluate the {\black contribution} of the weighting strategy.
This baseline MV REG is denoted as ``{nWSS}''.
In summary, we comprehensively compare the proposed WSS REG with five different MV REGs, including SSD, Boundary, Center, TV, and nWSS.
In addition, we replace the $\mathbf{S}_{\text{qu}}$ in \eqref{Q-norm REG} with the estimated abundance $\widehat{\mathbf{S}}$ from classical DIP, and maintain the same trade-off parameter $\lambda_2:=$1E-2 to validate the effectiveness of our QDIP.
Here, the widely {\black used} UnDIP is selected as the classical DIP.
{\black UnDIP first performs the geometry-based endmember extraction, followed by abundance estimation using a classical convolutional network.
Accordingly, the entire implicit model prior of the classical DIP exclusively influences the resulting abundance maps, which aligns with the proposed QDIP and enables a fair comparison in terms of abundance regularization.
More importantly, the quantitative performance of UnDIP is the second best in $\phi_{eb}$ and RMSE (cf. Table \ref{tab: performance}).}
This representative classical DIP will facilitate a persuasive ablation study.

Subsequently, we {\black present} details on the experimental settings of this ablation study {\black below}.
In the following experiment, we select one type of MV REG and DIP to regularize the proposed GQ-$\mu$ framework, {\black resulting in} 12 combinations (6 types of MV REGs and 2 types of DIP).
Due to space limitations, we {\black summarize} the quantitative results in Supplementary Table IV, where the {\black selected} MV REG/DIP will be marked by ``\CheckmarkBold''.
Regarding the {\black parameter} settings of the trade-off parameters for MV REGs, since the SSD, Boundary, Center, and TV REGs involve only a single term in their formulation, we {\black employ} the same $\lambda_3^{\dag}:=$1E-1 as the proposed GQ-$\mu$ for these MV REGs. 
For the nWSS and WSS REGs, we {\black retain} the same settings, $\lambda_3^{\dag}:=$1E-1 and $\lambda_4^{\dag}:=$1E-2, as presented in Section \ref{subsec:Experimental Protocol}.
All experiments {\black in} this ablation study are based on the {\black averaged} quantitative results (i.e., $\phi_{\text{en}}$, $\phi_{\text{ab}}$, and RMSE) across the three datasets (see Figure \ref{fig: ROI}).
Since our GQ-$\mu$ demonstrates stable results across runs (cf. Supplementary Table III), the quantitative results are reported from a single inference.

We now {\black turn to} the analysis.
First, we {\black employ} the classical DIP and compare the effectiveness of the MV REGs, as presented in the first six rows of Supplementary Table IV.
In these cases, we observe that the Boundary REG provides the most precise estimations for abundance and signatures.
By comparison, the proposed WSS REG {\black appears to} be effective only in abundance estimation with the classical DIP; and the rest of the MV REGs seem to achieve competitive performance in terms of $\phi_{\text{en}}$, $\phi_{\text{ab}}$, and RMSE.
Based on this observation, we experimentally assume that the proposed WSS REG can effectively help the NMF procedure with a strong abundance REG (e.g., QDIP) since the shrinkage weights are related to abundance.
Accordingly, we replace the classical DIP with the proposed QDIP and again evaluate the performance of each MV REGs.
As {\black demonstrated} in the last 6 rows of Supplementary Table IV, the results show that the quantitative metrics for abundance (i.e., $\phi_{\text{ab}}$ and RMSE) are significantly improved across all MV REGs.
These improvements greatly substantiate the effectiveness and necessity of the proposed QDIP.
Moreover, the proposed GQ-$\mu$ (with WSS REG and QIDP) further achieves remarkable improvements in both abundance and signature estimations.
The quantitative performance substantially outperforms the other 11 cases.
These experiments have accordingly demonstrated the effectiveness and necessity of our WSS REG and QDIP.
\begin{figure}[t]
    \centering
    \includegraphics[width=1\linewidth]{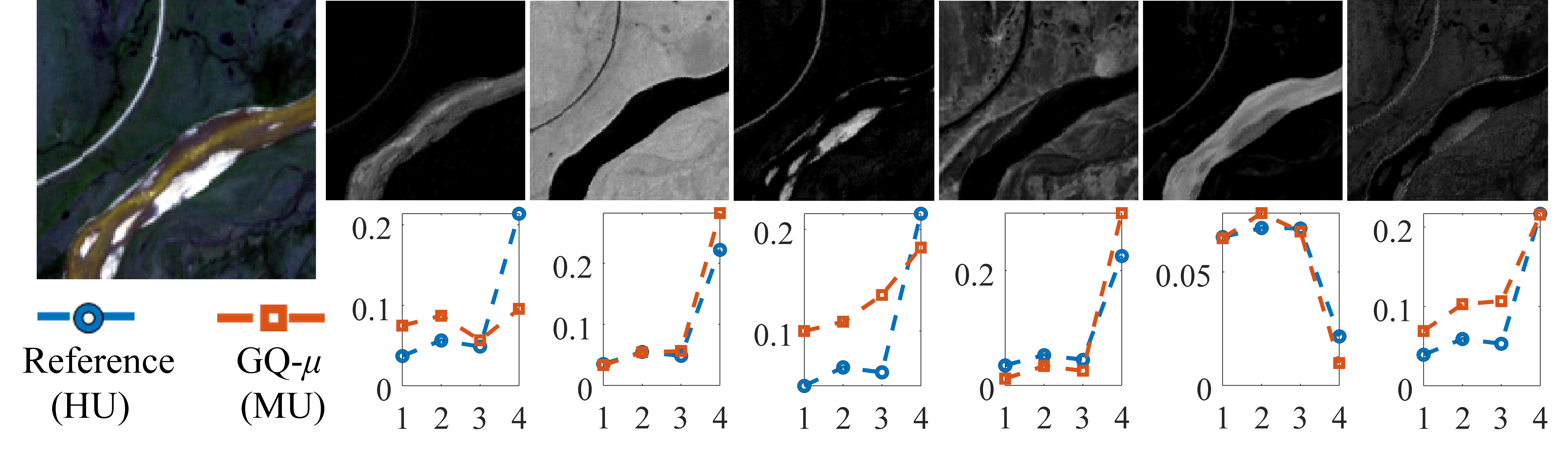}
    \caption{True-color composition (left) of the real Sentinel-2 MSI acquired at Northern Alaska, United States; and the 6 estimated multispectral endmembers and their corresponding abundances (right) using the proposed GQ-$\mu$.}
    \label{fig: Real_data_1}
\end{figure}
\begin{figure}[t]
    \centering
    \includegraphics[width=1\linewidth]{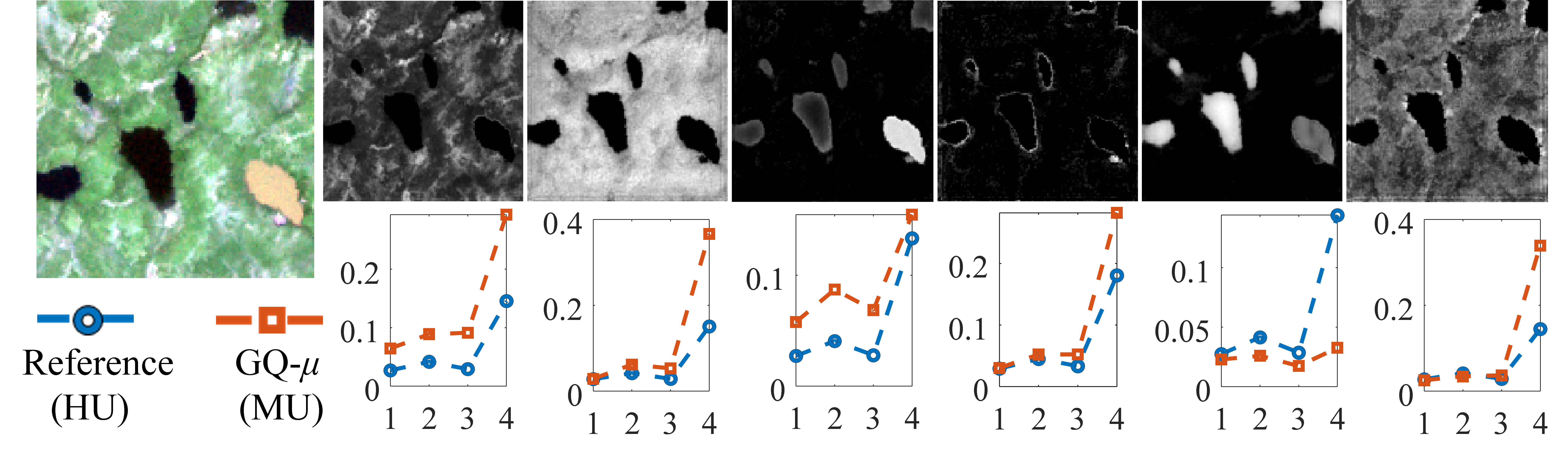}
    \caption{True-color composition (left) of the real Sentinel-2 MSI acquired at Mackenzie River Delta, Northwest Territories, Canada; and the 6 estimated multispectral endmembers and their corresponding abundances (right) using the proposed GQ-$\mu$.}
    \label{fig: Real_data_2}
\end{figure}

\subsection{Real-World MU for Sentinel-2 Data}\label{subsec: Real_data}
This section {\black presents} the real-data MU to demonstrate the {\black practical applicability} of our GQ-$\mu$. 
In remote sensing applications, multispectral sensors typically {\black acquire} blue (B), green (G), red (R), and near-infrared (NIR) bands (e.g., SPOT-7 \cite{SPOT7} and RapidEye \cite{RapidEye}), facilitating agricultural monitoring, land cover detection, and vegetation index estimation (e.g., NDVI).
In this analysis, we {\black utilize} MSIs acquired by Sentinel-2 imagery for {\black cross-dataset} validation.
Specifically, Sentinel-2 provides 12 bands at multiple resolutions, {\black where} the B, G, R, and NIR bands share the highest spatial resolution \cite{SSSS}.
We thus extract these bands as a 4-band MSI with $128\times128$ pixels and perform MU {\black with} the proposed GQ-$\mu$.
To {\black preserve} spatial consistency, we remove the upsample from the TConv Module 4 of QDIP (cf. Supplementary Table II), while retaining the other settings (cf. Section \ref{subsec:Experimental Protocol}) for the proposed GQ-$\mu$.
Owing to the lack of ground-truth abundances, the estimated abundance is rigorously evaluated with the high-resolution Google Earth imagery and relevant geomorphological references.
The {\black following} analysis demonstrates that all estimated abundances exhibit consistent spatial coherence with geomorphological patterns.  
Furthermore, we validate the estimated endmembers using full Sentinel-2 data and overdetermined endmember extraction (OEE).
First, the resolutions of 12 bands are unified by Sentinel-2 super-resolution via scene-adapted self-similarity (SSSS) \cite{SSSS}, followed by conducting the OEE using HyperCSI \cite{HyperCSI}.
{\black Subsequently}, the B, G, R, and NIR spectra are treated as reference endmembers, which correspond to ``Reference (HU)'' in Figure \ref{fig: Real_data_1} and Figure \ref{fig: Real_data_2}.
The first real dataset (cf. Figure \ref{fig: Real_data_1}) was acquired over Northern Alaska (NA), United States, in July 2019. 
The curvilinear feature in the upper-left region is the Dalton Highway, as identified {\black from} Google Earth imagery.
The curve of the lower-right area {\black corresponds to} the inland branch of Sagavanirktok River.
Besides, the distinct white blocks near the riverbank are potentially interpreted as the sedimentary packages \cite{gillis2014geologic}. 
{\black Most of} the land surface {\black in} this region is composed {\black primarily} of permanent permafrost, mineral, and vegetation of the tundra.
The second real dataset (cf. Figure \ref{fig: Real_data_2}) was {\black acquired} over Mackenzie River Delta (MRD), Northwest Territories, Canada, in July 2019. 
The MRD is notable for the largest concentration of pingos (i.e., small ice-cored hills formed in a {\black piecewise manner}) in the world \cite{burn2009mackenzie}.
In this ROI, the dark irregular patches correspond to open-water inland lakes, while the yellow patch is identified as an ice-covered water body according to the high-resolution Google Earth imagery. 
The surrounding uplands are ice-rich, and the unambiguous white patterns likely represent the ice-pure tabular bodies formed at the beginning of ground freezing.

The MU results for the NA dataset are {\black illustrated} in Figure \ref{fig: Real_data_1}.
From left to right, the 2nd, 3rd, and 5th abundances characterize the spatial distributions of the background land coverage, sedimentary packages, and the main component of the river, respectively.
The 4th and 6th abundances appear to {\black correspond} to mixed substances since this region is composed of multiple substances, including permanent permafrost, minerals, and vegetation of the tundra. 
We {\black further observe} the yellow curve in the center of the river (cf. the ROI in Figure \ref{fig: Real_data_1}); the corresponding proportion is more significant in the 1st abundance instead of the 5th abundance.
Although the 1st and 5th abundance seem similar, their endmembers are substantially distinct, {\black indicating} that they likely {\black correspond to} different materials.
{\black Moreover}, in the 5th abundance, the locations of sedimentary packages and the yellow curve exhibit lower proportions, consistent with the 1st and 5th abundance.
Since they share close locations (the reflectance may be affected by the water body), it is reasonable that these locations {\black retain} a moderate proportion in the 5th abundance.
Additionally, the multispectral endmembers estimated by our GQ-$\mu$ are {\black well aligned} with the reference endmembers, implying the precise MU results of GQ-$\mu$.
The MU results for the MRD dataset are presented in Figure \ref{fig: Real_data_2}.
The 1st and 2nd abundances characterize the spatial distribution of ice-pure tabular bodies and the background grounding, respectively.
The 3rd, 4th, and 5th abundances correspond to the open-water lakes, edges of water bodies, and the ice-covered lakes, respectively.
In the abundance of open water, the locations at ice-covered lakes retain low proportions since the ice coverages are floating on water.
The estimated endmembers from the proposed GQ-$\mu$ are again consistent with those obtained by OEE. 
The above real MU analysis further supports the effectiveness of the proposed GQ-$\mu$ for the challenging MU.

{\black The MRD dataset is further used to investigate the multispectral identifiability.
Specifically, the 4-band raw multispectral data is classified by K-means clustering with the clustering centers fixed as the MU-based endmember signatures (i.e., the orange curves in Figure \ref{fig: Real_data_2}).
As shown in Supplementary Figure 6, the clustering results do exhibit a strong spatial coherence with the MRD dataset, alluding that the 4-band multispectral data still hold material identifiability to some extent.
Nevertheless, in Supplementary Figure 6, the ice-covered water body (cf. the red box region) and the open-water inland lakes (cf. the blue box regions) are classified into the same cluster, mainly owing to the limited multispectral representation.
In contrast, GQ-$\mu$ successfully separates the ice region and open-water lakes into two clearly distinguishable abundance maps (cf. Figure \ref{fig: Real_data_2} and Supplementary Figure 6).
In other words, while the multispectral data alone may provide preliminary material identifiability, our GQ-$\mu$ algorithm further improves the discriminability of homogeneous materials.}

\section{Conclusions and Future Works}\label{sec:conclusion}
As current optical satellites mostly acquire MSIs with limited spatial resolution, multispectral unmixing (MU) {\black has become} a critical signal processing technology.
Hence, this work develops an unsupervised geometry/quantum-empowered MU (GQ-$\mu$) algorithm for solving this underdetermined blind source separation problem (i.e., MU).
{\black We first compute the counterpart HSI from a given MSI to transform the problem into a virtual overdetermined hyperspectral domain via the unsupervised band-splitting (BSP) function.
In the BSP [cf. \eqref{equ:LS}], the spectral deviation variable for each splitting band plays a critical role.
When it is set to excessively large, the resulting splitting bands may extend beyond their wavelength ranges.
Conversely, an overly small deviation hardly provides sufficient spectral variation and hence fails to yield informative structures.
Therefore, the deviation needs to be band-adaptive. 
In our work, we find that setting the deviation as one quarter of the difference between two adjacent bands yields promising results.
On the other hand,} we invent the first quantum version of the seminal deep image prior (DIP) {\black technique} to extract specific quantum features to regularize the abundance solution, and tailor a new simplex-geometry regularizer to estimate the hyperspectral endmembers, which are then spectrally downsampled back to the MSI space to obtain the multispectral endmembers.
Ablation study also validates the mechanics-inspired WSS geometry regularizer.
The proposed quantum DIP has mathematically guaranteed FE (ensuring a light-weight QUEEN), and ablation study demonstrates {\black the strength of} quantum DIP (not achieved by classical DIP).
All the above mechanisms can be done under a fully \textit{blind} setting.

In the future, we may {\black further} advance the unsupervised spectral augmentation, and hence facilitate a more precise endmember/abundance estimation in MU.
In the proposed GQ-$\mu$, the BSP function is designed to provide an initialization of virtual HSI.
Then, the proximal mapping, implemented by DnCNN denoising (originally for grayscale images), refines its spatial structure to {\black compute} the virtual HSI.
A natural improvement is to consider the spectral-spatial structure simultaneously, such as incorporating spectral regularization or a spatial-spectral denoiser.
Furthermore, inspired by the basis-learning strategy \cite{SSSS}, designing a customized update strategy for the virtual HSI during AO can potentially yield a better endmember/abundance estimation.
Regarding the complexity of QUEEN, a feasible direction is to explore a new unsupervised or weakly-supervised strategy for pretraining.
Once trained, the quantum inference can be formulated as a multiplication of structural matrices (e.g., sparse or unitary) \cite{HyperQUEEN}, offering a {\black practical strategy} for hardware acceleration and efficient quantum gate deployment.

\renewcommand{\thesubsection}{\Alph{subsection}}
\bibliography{ref_mu}

\vspace{-0.8cm}
\begin{IEEEbiography}[{\resizebox{0.9in}{!}{\includegraphics[width=1in,height=1.25in,clip,keepaspectratio]{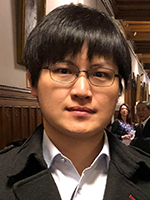}}}]
	{\bf Chia-Hsiang Lin}
(S'10-M'18-SM'24)
received the B.S. degree in electrical engineering and the Ph.D. degree in communications engineering from National Tsing Hua University (NTHU), Taiwan, in 2010 and 2016, respectively.
From 2015 to 2016, he was a Visiting Student of Virginia Tech,
Arlington, VA, USA.

He is currently a Professor with the Department of Electrical Engineering,
National Cheng Kung University (NCKU), Taiwan.
Before joining NCKU, he held research positions with The Chinese University of Hong Kong, HK (2014 and 2017),
NTHU (2016-2017),
and the University of Lisbon (ULisboa), Lisbon, Portugal (2017-2018).
He was an Assistant Professor with the Center for Space and Remote Sensing Research, National Central University, Taiwan, in 2018, and a Visiting Professor with ULisboa, in 2019.
His research interests include network science,
quantum computing,
convex geometry and optimization, blind signal processing, and imaging science.

Dr. Lin received the Emerging Young Scholar Award (The 2030 Cross-Generation Program) from National Science and Technology Council (NSTC), from 2023 to 2027,
the Future Technology Award from NSTC, in 2022,
the Outstanding Youth Electrical Engineer Award from The Chinese Institute of Electrical Engineering (CIEE), in 2022,
the Best Young Professional Member Award from IEEE Tainan Section, in 2021,
the Prize Paper Award from IEEE Geoscience and Remote Sensing Society (GRS-S), in 2020,
the Top Performance Award from Social Media Prediction Challenge at ACM Multimedia, in 2020,
and The 3rd Place from AIM Real World Super-Resolution Challenge at IEEE International Conference on Computer Vision (ICCV), in 2019.
He received the Ministry of Science and Technology (MOST) Young Scholar Fellowship, together with the EINSTEIN Grant Award, from 2018 to 2023.
In 2016, he was a recipient of the Outstanding Doctoral Dissertation Award from the Chinese Image Processing and Pattern Recognition Society and the Best Doctoral Dissertation Award from the IEEE GRS-S.
\end{IEEEbiography}
\vspace{-0.6cm}
\begin{IEEEbiography}[{\resizebox{0.9in}{!}{\includegraphics[width=1in,height=1.25in,clip,keepaspectratio]{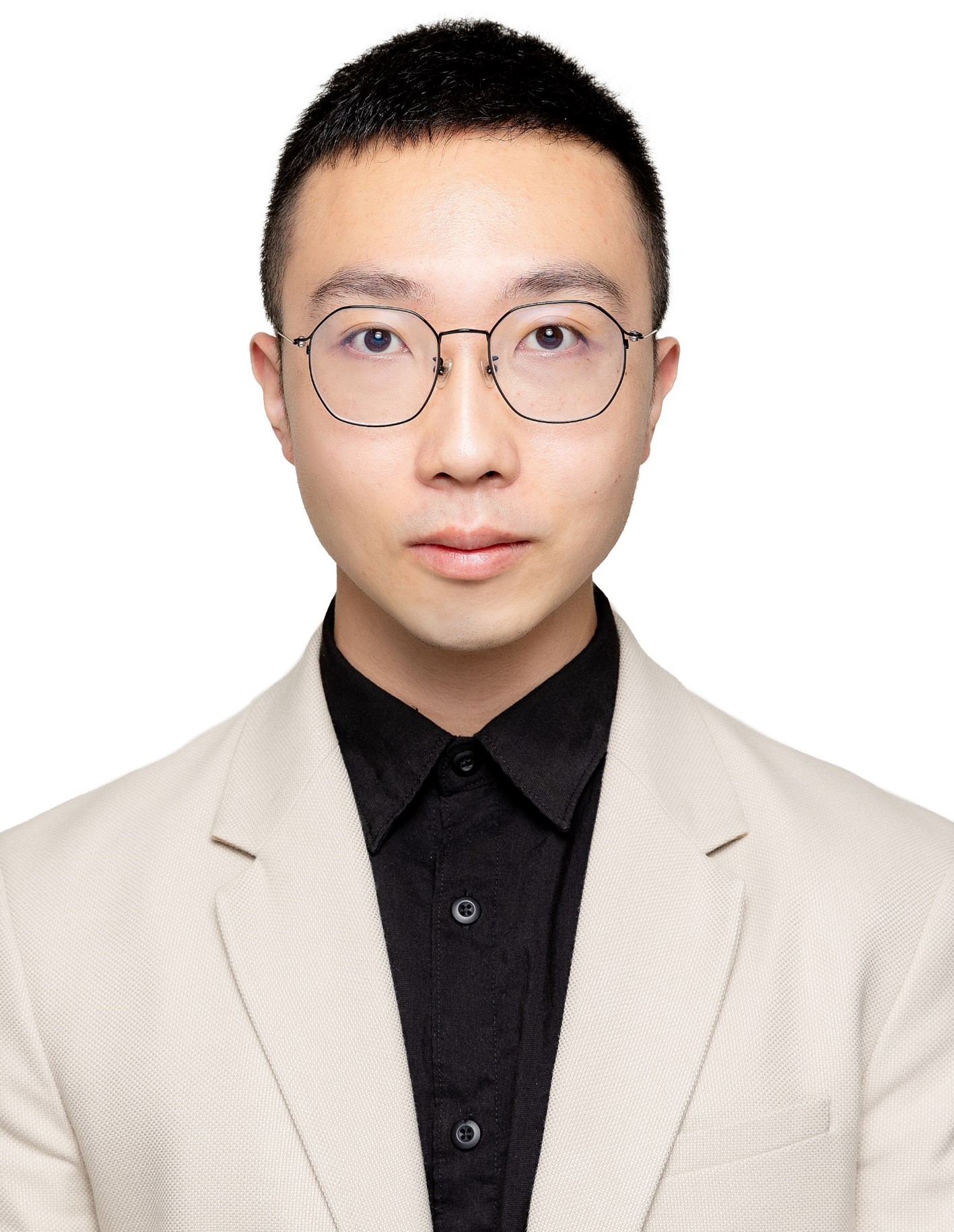}}}]
	{\bf Si-Sheng Young}
(S'23) is currently a Ph.D. student with the Intelligent
Hyperspectral Computing Laboratory (IHCL), National Cheng Kung University (NCKU), Tainan, Taiwan. 

In 2023 to 2024, he received the Merit Award from The Grand Challenge ``Computing for the Future", Miin Wu School of Computing, NCKU, ``Pan Wen Yuan Scholarship" from Industrial Technology Research Institute, Hsinchu, Taiwan, and ``Scholarship Pilot Program to Cultivate Outstanding Doctoral Students" from National Science and Technology Council (NSTC), Taipei, Taiwan.
His research interests include convex optimization, deep learning, anomaly detection, and imaging inverse problems.
\end{IEEEbiography}
\end{document}